\documentclass[11pt,a4paper]{article}
\usepackage{jheppub} 
\usepackage{epsfig,slashed}
\usepackage[utf8]{inputenc}
\usepackage{tensor}
\usepackage{tipa}
\usepackage{textcomp}
\usepackage{latexsym}
\usepackage{amsfonts}
\usepackage{array}
\usepackage{bigstrut}
\usepackage{bm,bbold}
\def\!{\mskip-\thinmuskip}

\newcommand{\mean}[1]{\langle#1\rangle}

\newcommand{\di}{{\rm d}}
\newcommand{\D}{{\rm d}}
\newcommand{\Dpi}{\mathcal{D}}
\newcommand{\mc}{\ensuremath{\mathcal}}  
\newcommand{\ii}{ \mathrm{i}}
\newcommand{\I}{ \mathrm{i}}
\newcommand{\h}[1]{\widehat{#1}}
\newcommand{\tr}{{\rm tr}}
\newcommand{\e}{{\rm e}}
\newcommand{\E}{{\rm e}}
\newcommand{\de}{\partial}
\newcommand{\ped}[1]{_{\textup{#1}}}
\newcommand{\apic}[1]{^{\textup{#1}}}

\newcommand{\group}[1]{\relax\ifmmode\mathsf{#1}\else\textsf{#1}\fi}
\renewcommand{\vec}[1]{\ensuremath{\mathchoice
		{\mbox{\boldmath$\displaystyle\mathbf{#1}$}}
		{\mbox{\boldmath$\textstyle\mathbf{#1}$}}
		{\mbox{\boldmath$\scriptstyle\mathbf{#1}$}}
		{\mbox{\boldmath$\scriptscriptstyle\mathbf{#1}$}}}}

\def\wT{{\widehat T}}
\def\wj{{\widehat j}}
\def\wJ{{\widehat J}}
\def\wK{{\widehat K}}
\def\wP{{\widehat P}}
\def\wQ{{\widehat Q}}

\def\wPsi{{\sf{\Psi}}}

\def\wrho{{\widehat{\rho}}}
\newcommand{\oper}{\mathcal{O}_{\alpha\dots\beta}}

\newcommand{\oraw}[1]{\overrightarrow{#1}}
\newcommand{\olaw}[1]{\overleftarrow{#1}}

\DeclareMathOperator*{\sumint}{
	\mathchoice%
	{\ooalign{$\displaystyle\sum$\cr\hidewidth$\displaystyle\int$\hidewidth\cr}}
	{\ooalign{\raisebox{.14\height}{\scalebox{.7}{$\textstyle\sum$}}\cr\hidewidth$\textstyle\int$\hidewidth\cr}}
	{\ooalign{\raisebox{.2\height}{\scalebox{.6}{$\scriptstyle\sum$}}\cr$\scriptstyle\int$\cr}}
	{\ooalign{\raisebox{.2\height}{\scalebox{.6}{$\scriptstyle\sum$}}\cr$\scriptstyle\int$\cr}}
}

\newcommand{\mycorr}[1]{\langle\langle#1\rangle\rangle}

\hypersetup{pdfinfo={Title={General thermodynamic equilibrium with chiral chemical potential},Author={Matteo Buzzegoli}}}

\begin{document}

\title{General thermodynamic equilibrium with axial chemical potential for the free Dirac
field} 

\author[a]{M. Buzzegoli}
\author[a]{and F. Becattini}
\affiliation[a]{Universit\`a di Firenze and INFN Sezione di Firenze,\\
Via G. Sansone 1, I-50019 Sesto Fiorentino (Firenze), Italy}

\emailAdd{matteo.buzzegoli@unifi.it}

\abstract{
We calculate the constitutive equations of the stress-energy tensor and the currents 
of the free massless Dirac field at thermodynamic equilibrium with acceleration and rotation and a 
conserved axial charge by using the density operator approach. We carry out an expansion 
in thermal vorticity to the second order with finite axial chemical potential $\mu_A$. 
The obtained coefficients of the expansion are expressed as correlators of angular
momenta and boost operators with the currents. We confirm previous observations that the 
axial chemical potential induces non-vanishing components of the stress-energy tensor 
at first order in thermal vorticity due to breaking of parity invariance of the density 
operator with $\mu_A \ne 0$. The appearance of these components might play an important 
role in chiral hydrodynamics.}
\maketitle
\flushbottom

\section{Introduction}

The physics of chiral fluids has drawn remarkable attention lately. Much of the interest is
driven by the quest of the Chiral Magnetic Effect (CME) in relativistic heavy ion 
collisions~\cite{Kharzeev:2015znc}. Besides, the discovery of a finite global polarization of the
$\Lambda$ hyperons at the Relativistic Heavy Ion Collider (RHIC)~\cite{STAR:2017ckg} 
in agreement with the predictions of local thermodynamic equilibrium \cite{Becattini:2007sr}
has demonstrated that the Quark-Gluon Plasma (QGP) has sufficient local vorticity and
acceleration to induce polarization of produced particles. Because of the relation between spin 
and chirality, this phenomenon ought to have a close connection with the physics of CME.

Most of the studies of the chiral fluids were aimed at determining both thermodynamic 
equilibrium and non-equilibrium coefficients (either dissipative or non-dissipative 
according to the classification of ref.~\cite{Haehl:2014zda}) within different theoretical
frameworks: in relativistic hydrodynamics~\cite{Neiman:2010zi,Pu:2010as};
with Kubo Formulae in holography~\cite{Landsteiner:2011iq,Landsteiner:2012kd};
in finite temperature field theory~\cite{Chowdhury:2015pba,Golkar:2012kb};
in chiral kinetic theory~\cite{Chen:2015gta,Hidaka:2017auj,Abbasi:2017tea,Huang:2018wdl};
and in effective field theory~\cite{Glorioso:2017lcn,Huang:2017pqe}.

Thermodynamic equilibrium coefficients include - amongst others - the well known proportionality 
coefficient between vector current and magnetic field in the CME, originally derived
in ref.~\cite{Fukushima:2008xe} assuming equilibrium. In fact, the systematic
derivations of all constitutive relations at thermodynamic equilibrium, which is non-dissipative
by construction, can be carried out with a functional approach \cite{Kovtun2017,Kovtun2018},
with finite temperature field thoery~\cite{Ambrus:2014uqa}
or with the covariant density operator approach~\cite{Becattini:2014yxa}. The latter was used to derive the constitutive relations of the stress-energy tensor in presence of acceleration and vorticity.

In this paper, we will take this approach to study the thermodynamic equilibrium of a chiral
fluid endowed with an axial chemical potential (i.e. a non-vanishing mean axial charge) including
acceleration and vorticity and we will present explicit calculation for the free massless Dirac
field. We will see that the presence of a non-vanishing axial charge is responsible for additional 
parity-breaking terms in the constitutive equations of the stress-energy tensor and currents,
which would not be otherwise present. These terms can be very important for the physics of CME
and particularly for the the implementations of chiral magnetohydrodynamics. Indeed, general thermodynamic 
equilibrium relations are the fundamental building blocks of hydrodynamics once they are 
promoted to space-time dependent functions. Particularly, we will see that the axial chemical
potential entails the appearance of new terms which are first-order in the gradient expansion.

\subsection*{Notation}

In this paper, we use the natural units, with $\hbar=c=k_B=1$.
The Minkowskian metric tensor is ${\rm diag}(1,-1,-1,-1)$; for the Levi-Civita symbol we use the 
convention $\epsilon^{0123}=1$. Operators in Hilbert space will be denoted by a large upper hat, 
e.g. $\wT$ (with the exception of Dirac field operator that is denoted
by $\wPsi$). The stress-energy tensor used to define Poincaré generators is always assumed to 
be symmetric with an associated vanishing spin tensor.

\section{Thermodynamic equilibrium and chirality}

In this section we will summarize the density operator approach and the inclusion of
a conserved axial charge. 

The covariant density operator formalism has been described in detail elsewhere, here
we will briefly summarize it. The most general equilibrium density operator can be obtained 
by maximizing the total entropy $S=-\tr(\h\rho\log\h\rho\,)$ with the constraints of given 
energy, momentum and charge densities over some spacelike hypersurface $\Sigma$~\cite{weert,Becattini:2014yxa,Hayata:2015lga} and can be written as~\cite{Zubarev,weert,Becattini:2012tc}, see also~\cite{Becattini:2015nva,Hongo:2016mqm},
\begin{equation}
\label{ltematrix}
\wrho=\frac{1}{\mc Z}\exp\left[-\int_\Sigma \D\Sigma_\mu \left(\wT^{\mu\nu}(x)\beta_\nu(x)-\sum_i\zeta_i(x)\wj^\mu_i(x)\right)\right],
\end{equation}
where  $\wT^{\mu\nu}$ and $\wj^\mu_i$ are the symmetric stress-energy tensor and some conserved currents
operators, $\beta^\mu$ is the four-temperature vector such that $T=1/\sqrt{\beta^2}$ 
is the proper comoving temperature and $\zeta_i$ are the ratio of comoving chemical potentials and
the temperature $\zeta_i=\mu_i/T$. The Eq.~(\ref{ltematrix}) describes a \emph{local} thermal equilibrium;
to obtain a \emph{global} equilibrium distribution the operator (\ref{ltematrix}) is required to be time independent, implying that the integrand is divergenceless. This is possible 
if~\cite{Becattini:2012tc} all $\zeta_i$ are constants and the four-temperature $\beta$
is a Killing vector field:
\begin{equation}
\label{betakilling}
\nabla_\mu\zeta_i=0,\qquad\nabla_\mu\beta_\nu+\nabla_\nu\beta_\mu=0.
\end{equation}
The general solution of Eq.~(\ref{betakilling}) in Minkowsky space-time is
\begin{equation}
\label{betaminkowski}
	\zeta_i=\text{constant,}\qquad\beta_\mu=b_\mu+\varpi_{\mu\nu}x^\nu,
\end{equation}
where $b$ is a constant four-vector and $\varpi$ a constant antisymmetric tensor, dubbed \emph{thermal vorticity},
that accounting for Eq. (\ref{betakilling}) can be expressed as an exterior derivative of the $\beta$ field
\begin{equation}\label{thvort}
	\varpi_{\mu\nu}=-\frac{1}{2}\left(\de_\mu\beta_\nu-\de_\nu\beta_\mu \right).
\end{equation}
Recalling the definition of four-momentum operator $\h P$, conserved charges $\h Q_i$
\begin{equation*}
\wP^\mu=\int_\Sigma \D\Sigma_\lambda\, \wT^{\lambda\mu},\qquad
\wQ_i=\int_\Sigma \D\Sigma_\lambda\, \wj^\lambda_i
\end{equation*}
and generators of Lorentz transformation $\wJ$
\begin{equation}
\label{Lorentzgen}
\wJ^{\mu\nu}=\int_\Sigma \D\Sigma_\lambda \left(x^\mu\wT^{\lambda\nu}-x^\nu\wT^{\lambda\mu}\right),
\end{equation}
the form (\ref{betaminkowski}) allows to write the density operator (\ref{ltematrix}) at global equilibrium as
\begin{equation*}
\wrho=\frac{1}{\mc Z}\exp\Big[ -b_\mu\wP^\mu +\frac{1}{2}\varpi_{\mu\nu}\wJ^{\mu\nu}+\sum_i\zeta_i\,\wQ_i\Big].
\end{equation*}
The state characterized by this density operator is a generalization of global equilibrium, and since the parameters are globally defined we will refer to this condition as \emph{general global equilibrium}.

As conserved currents, one usually have electric and baryon currents. However, for massless 
fermions there is one additional conserved current, the axial current $\wj_A$, such that:
\begin{equation}\label{axcurcons}
\nabla_\mu \wj_A^{\,\mu} = 0
\end{equation}
which has an associated conserved axial charge
\begin{equation}\label{ConsCharges}
\wQ_A=\int_\Sigma \D\Sigma_\lambda\, \wj^\lambda\ped{A}
\end{equation}
where $\Sigma$ is an arbitrary space-like hypersurface. Thus, for a system of massless fermions, one would 
deal with a general global thermodynamic equilibrium of this sort:
\begin{equation}
\label{chiralstat}
\wrho\ped{GE}=\frac{1}{\mc{Z}\ped{GE}}\exp\Big[ -b_\mu\wP^\mu +\frac{1}{2}\varpi_{\mu\nu}\wJ^{\mu\nu}+\zeta\,\wQ+\zeta\ped{A}\,\wQ\ped{A}\Big].
\end{equation}
In fact, the axial current is not conserved in presence of interactions because of the well known 
axial anomaly and the above continuity equation gets modified into (for QED):
\begin{equation}\label{anomaly}
 \nabla_\mu \wj_A = C F^{\mu\nu} \tilde F_{\mu\nu}\, .
\end{equation}
Nevertheless, in presence of an external field such that the right hand side of (\ref{anomaly})
vanishes, the axial charge is still constant and thermodynamic equilibrium is possible. 
If, on the other hand, we deal with dynamical gauge fields, the right hand side can be written 
as the divergence of the so-called Chern-Simons current:
\begin{equation*}
 K^\mu = C \,\epsilon^{\mu\nu\rho\sigma} A_\nu F_{\rho\sigma}
\end{equation*}
and its integral (the helicity, for the EM field) can be embodied in the definition 
of a gauge-invariant overall conserved axial charge including both fermion and gauge boson 
contribution~\cite{Boyarsky:2015faa}. In both the above cases, the density operator (\ref{chiralstat}) is perfectly
meaningful.

In this work, we will confine ourselves to the simplest case of free fermions, so that
(\ref{axcurcons}) is fulfilled. Thus, the obtained relations will not include contributions 
from possible external gauge fields, yet they will be very useful as a leading order 
expressions for a chiral fluid.

As a final remark, it is important to stress that $\beta$ must be a time-like vector 
for the local mean values to be physical (see next Section). In general, for $\varpi\neq 0$ 
there is a 3D boundary in space-time separating a region where $\beta$ is space-like;
for an analysis of the effect of the boundary inside this region on rotating fermions
see~\cite{Ambrus:2015lfr}.
In the physical region, the four-temperature vector $\beta$ defines a four-velocity $u$ 
of the fluid - the $\beta$ \emph{frame}~\cite{Becattini:2014yxa} or thermodynamic 
frame~\cite{Kovtun2017} -
\begin{equation}
\label{fluidvelocity}
u^\mu(x)=\frac{\beta^\mu(x)}{\sqrt{\beta^2}}.
\end{equation}
As we will see, this frame is the most appropriate for thermodynamic equilibrium.

\section{Mean value of local operators}

Our goal is to calculate the mean values of physical quantities at thermodynamic equilibrium
with the density operator $\wrho$ in Eq.~(\ref{chiralstat}):
\begin{equation*}
\mean{\h{O}(x)}=\tr\left(\wrho\,\h{O}(x) \right)\ped{ren}.
\end{equation*}
For this purpose, we will use the method already employed 
in refs.~\cite{Becattini:2015nva,Buzzegoli:2017cqy}.

The first step is to shift the angular momentum operator at $x$ by using the translational operator 
$\h {\sf T}(x)=\exp[\,\I x\cdot\wP\,]$:
\begin{equation}\label{jshift}
\begin{split}
\h J^{\mu\nu}_x \equiv&  \h {\sf T}(x) \h J^{\mu\nu} \h {\sf T}^{-1}(x)=
\h J^{\mu\nu}-x^{\mu}\h P^{\nu}+x^{\nu}\h P^{\mu}\\
&=\int_\Sigma \di \Sigma_{\lambda} \left[ (y^\mu -x^{\mu})\h 
T^{\lambda\nu}(y)-(y^\nu - x^{\nu})\h T^{\lambda\mu}(y)\right].
\end{split}
\end{equation}
Thanks to this definition and Eq.~(\ref{betaminkowski}) we can rewrite the density
operator~(\ref{chiralstat}) as
\begin{equation}\label{chiraldensop}
\wrho\ped{GE} =\frac{1}{\mc{Z}}\exp\Big[ -\beta(x)\cdot\wP +\frac{1}{2}\varpi:\wJ_x+\zeta\,
\wQ+\zeta\ped{A}\,\wQ\ped{A}\Big].
\end{equation}

The second step is to make an expansion in thermal vorticity. Indeed, thermal vorticity~(\ref{thvort})
is adimensional in natural units and, under most circumstances (e.g. in 
relativistic heavy ion collisions) is $\ll 1$. An expansion of (\ref{chiraldensop}) in $\varpi$
will result in a good approximation of the mean values. At the leading order one has:
\begin{equation*}
\begin{split}
\mean{\h{O}(x)}\ped{GE}&\equiv\tr\left(\wrho\ped{GE}\,\h{O}(x) \right)
=\frac{1}{\mc{Z}\ped{GE}}\tr\left(\exp\Big[ -b_\mu\wP^\mu +\frac{1}{2}\varpi_{\mu\nu}\wJ^{\mu\nu}+\zeta\,\wQ+\zeta\ped{A}\,\wQ\ped{A}\Big]\,\h{O}(x) \right)\\
&\simeq \frac{1}{\mc{Z}_{\beta(x)}}\tr\left(\exp\Big[ -\beta(x)_\mu\wP^\mu +\zeta\,\wQ+\zeta\ped{A}\,\wQ\ped{A}\Big]\,\h{O}(x) \right)
\equiv \mean{\h{O}(x)}_{\beta(x)}=\mean{\h{O}(0)}_{\beta(x)},
\end{split}
\end{equation*}
where in the last step we have taken advantage of the translational invariance of the homogeneous
operator $ \propto \exp[-\beta \cdot \wP]$. The above expression shows that - as expected - for small
deviations from homogeneous equilibrium, the mean value is well approximated by the 
homogeneous equilibrium value with four-temperature equal to its value in the same point $x$
where the operator is evaluated.

The corrections to the leading order term are obtained by using the expansion based on a path 
ordered method developed in ref.~\cite{weert}, in our case along the fluid velocity $u$~(\ref{fluidvelocity}). This method is also appropriate for Euclidean space-time. The resulting expression is given 
in terms of connected correlators:
\begin{equation*}
\begin{split}
\mean{\h O}_{c}&=\mean{\h O}\\
\mean{\h J \h O}_{c}&= \mean{\h J\h O}-\mean{\h J}\mean{\h O}\\
\mean{\h J_1\h J_2\h O}_{c}&= \mean{\h J_1\h J_2\h O}-\mean{\h J_1}\mean{\h J_2 \h O}_{c}-
\mean{\h J_2}\mean{\h J_1 \h O}_{c}-\mean{\h O}\mean{\h J_1 \h J_2}_{c}-\mean{J_1}\mean{J_2}\mean{\h O}\\
&\vdots
\end{split}
\end{equation*}
weighted with the homogeneous statistical operator
\begin{equation}\label{leadorder}
\mean{\h O\,}_{\beta(x)}\equiv\frac{\tr\left[ \exp\left(-\beta(x)\cdot \wP+\zeta\wQ+\zeta\ped{A}\wQ\ped{A}\right)\h O\,\right]}{\tr\left[ \exp\left(-\beta(x)\cdot\wP+\zeta\wQ+\zeta\ped{A}\wQ\ped{A}\right)\right]}.
\end{equation}
At the second order in thermal vorticity we obtain~\cite{Buzzegoli:2017cqy}
\begin{equation}
\begin{split}\label{tauexp2}
\mean {\h O(x)}=&\mean{\h O(0)}_{\beta(x)}+\frac{\varpi_{\mu\nu}}{2|\beta|}\int_0^{|\beta|}\di \tau 
\mean{{\rm T}_\tau\left(\h J^{\mu\nu}_{-\ii \tau u}\h O(0)\right)}_{\beta(x),c}
\\& +\frac{\varpi_{\mu\nu}\varpi_{\rho\sigma}}{8|\beta|^2} \int_0^{|\beta|}\di \tau_1\di \tau_2
\mean{{\rm T}_\tau\left(\h J^{\mu\nu}_{-\ii \tau_1 u} \h J^{\rho\sigma}_{-\ii \tau_2 u}\h O(0)\right)}_{\beta(x),c}+\mathcal{O}(\varpi^3),
\end{split}
\end{equation}
where ${\rm T}_\tau\left(\cdots\right)$ is the ${\rm T}_\tau$-ordered product defined as
\begin{equation*}
{\rm T}_\tau \big(\h{O}_1(\tau_1) \h{O}_2(\tau_2) \cdots \h{O}_N(\tau_N)\big)
\equiv \h{O}_{p_1}(\tau_{p_1}) \h{O}_{p_2}(\tau_{p_2}) \cdots \h{O}_{p_N}(\tau_{p_N})
\end{equation*}
with $p$ the permutation that orders $\tau$ by value:
\begin{align*}
&p \mathrel{:} \{1, 2, \dots, N\} \to \{1, 2, \dots, N\}\\
&\tau_{p_1} \leq \tau_{p_2} \leq \cdots \leq \tau_{p_N}.
\end{align*}
%

Thermal vorticity can be decomposed into two space-like fields $\alpha$ and $w$ \cite{Becattini:2015nva}
much the same way as the electro-magnetic field tensor, by using the fluid four-velocity $u$:
\begin{equation*}
\varpi_{\mu\nu}=\epsilon_{\mu\nu\rho\sigma}w^\rho u^\sigma+\alpha_\mu u_\nu - \alpha_\nu u_\mu.
\end{equation*}
At global equilibrium, when $\beta$ is a Killing vector, it can be shown that $w$ and $\alpha$
are the vorticity and acceleration vectors divided by the local temperature $T$ \cite{Becattini:2015nva}:
\begin{equation}
\label{accvort}
w_\mu=-\frac{1}{2}\epsilon_{\mu\nu\rho\sigma}\varpi^{\nu\rho}u^\sigma=\frac{1}{T}\omega_\mu,\quad
\alpha_\mu=\varpi_{\mu\nu}u^\nu=\frac{1}{T}a_\mu.
\end{equation}
Furthermore, it is useful to define another four-vector:
\begin{equation}\label{transversedir}
\gamma_\mu=\epsilon_{\mu\nu\rho\sigma}w^\nu\alpha^\rho u^\sigma=(\alpha\cdot\varpi)^\lambda\Delta_{\lambda\mu},
\end{equation}
where $\Delta_{\mu\nu}=g_{\mu\nu}-u_\mu u_\nu$ is the projector onto the hyperplane perpendicular
to $u$. The four-vector $\gamma$ is second-order in thermal vorticity and is, by construction,
orthogonal to $u$, $w$ and $\alpha$. Similarly, the generators of the Lorentz group $\wJ$
can be decomposed as:
\begin{equation*}
\h{J}^{\mu\nu}=u^\mu\h{K}^\nu-u^\nu\h{K}^\mu-u_\rho\epsilon^{\rho\mu\nu\sigma}\h{J}_\sigma,
\end{equation*}
where
\begin{equation}\label{boostrot}
\h{K}^\mu=u_\lambda\h{J}^{\lambda\mu},\quad
\h{J}^\mu=\frac{1}{2}\epsilon^{\alpha\beta\gamma\mu}u_\alpha\h{J}_{\beta\gamma}\, .
\end{equation}
The operators $\wJ$ and $\wK$ are the generators of rotations and boosts with respect
to the local frame defined by $u$. Making use of~(\ref{accvort}), we obtain
\begin{align*}
\frac{1}{2}\varpi_{\mu\nu}\h{J}^{\mu\nu}&=-\alpha^\rho\h{K}_\rho-w^\rho\h{J}_\rho\\
\varpi_{\mu\nu}\varpi_{\rho\sigma}\h{J}^{\mu\nu}\h{J}^{\rho\sigma}&=
2\alpha^\mu\alpha^\nu \{\h{K}_\mu,\,\h{K}_\nu\}+2w^\mu w^\nu \{\h{J}_\mu,\,\h{J}_\nu\}
+4\alpha^\mu w^\nu \{\h{K}_\mu,\,\h{J}_\nu\}
\end{align*}
whence the mean value expansion~(\ref{tauexp2}) becomes
\begin{equation*}
\begin{split}
\mean {\h O(x)}=&\,\mean{\h O(0)}_{\beta(x)}
-\frac{\alpha_\rho}{|\beta|}\int_0^{|\beta|}\di \tau\, \mean{{\rm T}_\tau\left(\h K^{\rho}_{-\ii \tau u} \h O(0)\right)}_{\beta(x),c}\\
&-\frac{w_\rho}{|\beta|} \int_0^{|\beta|}\di \tau\, \mean{{\rm T}_\tau \left(\h J^{\rho}_{-\ii \tau u} \h O(0)\right)}_{\beta(x),c}\\
&+\frac{\alpha_\rho\alpha_\sigma}{2|\beta|^2}\int_0^{|\beta|}\di \tau_1\di \tau_2
\,\mean{{\rm T}_\tau \left(\h K^{\rho}_{-\ii \tau_1 u } \h K^{\sigma}_{- \ii \tau_2 u} \h O(0)\right)}_{\beta(x),c}\\
&+\frac{w_\rho w_\sigma}{2|\beta|^2}\int_0^{|\beta|}\di \tau_1\di \tau_2 \,\mean{{\rm T}_\tau
	\left(\h J^{\rho}_{-\ii \tau_1 u } \h J^{\sigma}_{-\ii \tau_2 u} \h O(0)\right)}_{\beta(x),c}\\
&+\frac{\alpha_\rho w_\sigma}{2|\beta|^2}\int_0^{|\beta|}\di \tau_1\di \tau_2
\,\mean{{\rm T}_\tau \left(\{\h K^\rho_{-\ii \tau_1 u},\h J^{\sigma}_{-\ii \tau_2 u}\}\h O(0)\right)}_{\beta(x),c}
+\mathcal{O}(\varpi^3)
\end{split}
\end{equation*}
or, after defining the correlators
\begin{equation}
\begin{split}\label{mycorr}
	\mycorr{\h K^{\rho_1}\cdots \h K^{\rho_n} \h J^{\sigma_1}\cdots \h J^{\sigma_m} \h O} \equiv&
	\int_0^{|\beta|} \frac{\di\tau_1\cdots\di\tau_{n+m}}{|\beta|^{n+m}}\times\\
	&\times\mean{{\rm T}_\tau\left(\h K^{\rho_1}_{-\ii \tau_1 u}\cdots\h K^{\rho_n}_{-\ii \tau_n u}
		\h J^{\sigma_1}_{-\ii \tau_{n+1} u} \cdots\h J^{\sigma_m}_{-\ii \tau_{n+m} u} \h O(0)\right)}_{\beta(x),c}
\end{split}
\end{equation}
the expansion in thermal vorticity of the mean value of a local operator $\widehat O(x)$
can be written as:
\begin{equation}
\begin{split} \label{meanvalueoper}
\mean {\h O(x)}=&\mean{\h O(x)}_{\beta(x)}-\alpha_\rho \mycorr{\,\h K^\rho \h O\,}-w_\rho \mycorr{\,\h J^\rho \h O\,}
+\frac{\alpha_\rho\alpha_\sigma}{2}\mycorr{\,\h K^\rho \h K^\sigma \h O\,}\\
&+\frac{w_\rho w_\sigma}{2} \mycorr{\,\h J^\rho \h J^\sigma \h O\,} +\frac{\alpha_\rho w_\sigma}{2}\mycorr{\,\{\h K^\rho,\h J^\sigma\}\h O\,}+\mathcal{O}(\varpi^3).
\end{split}
\end{equation}
Note that the above expansion applies to any local operator $\h O$, whether it is a scalar or
a component of a tensor of any rank.

\section{Chiral thermodynamic coefficients}

In this section, we identify and determine the thermodynamic coefficients up to second order 
in thermal vorticity for the stress energy tensor as well as vector and axial currents.
We have seen in the previous section, that the mean values of local observables can be expanded 
in thermal vorticity~(\ref{meanvalueoper}) and that each term involves a coefficient expressed
as connected correlator between the observables and the Lorentz group generators~(\ref{mycorr}),
caluculated with the homogeneous statistical operator $\wrho\ped{h}$~(\ref{leadorder})
\begin{equation}\label{hdensop}
\wrho\ped{h}=\frac{1}{{\mc Z}\ped{h}} \exp\left(-\beta(x)\cdot \wP+\zeta\wQ+\zeta\ped{A}\wQ\ped{A}\right).
\end{equation}
This operator features several symmetries: translations, rotations in the hyperplane perpendicular
to $\beta(x)$, and time-reversal $\group{T}$ in the $\beta(x)$ rest frame. However, as $\wQ$ 
is odd under charge conjugation $\group{C}$ and $\wQ_A$ is odd under parity $\group{P}$ (see Table~\ref{tab:TransfProp}), $\wrho\ped{h}$ is not invariant under $\group{C}$ and $\group{P}$. 

The symmetries of the density operator have direct consequences on the value of the aforementioned 
correlators and, more generally, on mean values of operators. For instance, time-reversal invariance 
of $\wrho\ped{h}$ implies that any mean value of $\mean{\h O(x)}_{\beta(x)}$, such as the ones
in~(\ref{mycorr}), vanishes if the operator $\h O(x)$ is odd under time-reversal:
$$
\group{T}\h O\group{T}=-\h O .
$$
Similarly, rotational invariance implies that correlators in~(\ref{mycorr}) will be non-vanishing
only if the involved operators transform like the same irreducible vectors of the same representations 
of the rotation group. To study the impact of symmetries on the correlators~(\ref{mycorr}) it 
is convenient to redefine: 
$$
\h K^{\rho_1}\cdots \h K^{\rho_n} \h J^{\sigma_1}\cdots \h J^{\sigma_m} = \h A
$$
and to decompose both $\h A$ and $\h O$ in irreducible components under rotation: scalar, vector,
symmetric traceless tensor, etc. The only non-vanishing correlators will be those matching corresponding
components of the same kind {\em and} with the same transformation property under time-reversal.

In the homogeneous equilibrium density operator (\ref{hdensop}), the term $\mu_A \wQ_A$ is 
the only responsible for parity breaking, if the Hamiltionian is parity invariant. For this
reason, all correlators~(\ref{mycorr}) in which $\h A$ and $\h O$ transform with different 
sign under parity will be proportional to odd powers of $\mu_A$, because:
$$
 \tr (\,\wrho_h\, \h A\, \h O\,) = \tr (\,\group{P}\, \wrho_h\, \h A\, \h O\, \group{P}^{-1}) = 
  \tr (\,\wrho_h(-\mu_A) \group{P} \h A\, \h O\, \group{P}^{-1}) = - \tr (\,\wrho_h(-\mu_A) \h A\, \h O\,).
$$
These correlators will be henceforth dubbed as \emph{chiral correlators}, and the corresponding
coefficients chiral coefficients (see Table~\ref{tab:TransfProp}). 

\begin{table}[htb]
	\caption{Transformation properties under discrete transformation for various operators in the rest frame.
		$\group{P}$ is parity, $\group{T}$ is time-reversal and $\group{C}$ is charge conjugation transformation.
		$\wT^{\mu\nu}$ is the stress-energy tensor, $\wj\ped{V}^\mu$ is an (electric) vector current,  $\wj\ped{A}^\mu$ is an axial current,
		$\phi$ is a scalar and $\phi\ped{A}$ is a pseudoscalar,  $\wK^\mu$ and $\wJ^\mu$ are the Lorentz generators,
		$\wQ$ and $\wQ\ped{A}$ are vector and axial charge respectively.}
	\label{tab:TransfProp}
	\[\begin{array}{lccc|cccccc|cccc|cc}
	& \wT^{00} & \wT^{0i} & \wT^{ij} & \wj^0\ped{V} & \h{\vec{j}}\ped{V} & \wj^0\ped{A} & \h{\vec{j}}\ped{A} & \h\phi & \h\phi\ped{A} &
	\h{\vec{K}} & \h{\vec{J}} & (\wK\wK, \wJ\wJ) & \wK\wJ & \wQ & \wQ\ped{A}\\  
	\hline
	\group{P} & + & - & + & + & - & - & + & + & - & - & + & + & - & + & - \\
	\group{T} & + & - & + & + & - & + & - & + & - & + & - & + & - & + & + \\
	\group{C} & + & + & + & - & - & + & + & + & + & + & + & + & + & - & + \\
	\end{array}\]	
\end{table}

We refer to \cite{Becattini:2015nva,Buzzegoli:2017cqy} for a detailed description of the method
of calculating the coefficients; herein we briefly summarize it. 
The coefficients can be expressed as thermal expectation values
in the rest frame where $\beta^\mu=(1/T(x),\vec{0})$ and they are function of $T$, $\zeta = \mu/T$ and 
$\zeta_A = \mu_A/T$; we denote the mean values in rest frame with a subscript $T$, that is: 
\begin{equation*}
\mean{\h O(x)}_T\equiv\frac{\tr\left[ \exp\left(-\h H/T +\zeta\wQ+\zeta\ped{A}\wQ\ped{A}\right)\h O(0)\right]}{\tr\left[ \exp\left(-\h H/T +\zeta\wQ+\zeta\ped{A}\wQ\ped{A}\right)\right]},
\end{equation*}
where $\h H$ is the Hamiltonian.
Consequently, the correlators~(\ref{mycorr}) in the rest frame are written as:
\begin{equation}
\begin{split}\label{mycorrrest}
\mycorr{\h K^{\rho_1}\cdots \h K^{\rho_n} \h J^{\sigma_1}\cdots \h J^{\sigma_m} \h O}_T \equiv&
\int_0^{|\beta|=1/T} \frac{\di\tau_1\cdots\di\tau_{n+m}}{|\beta|^{n+m}}\times\\
&\times\mean{{\rm T}_\tau\left(\h K^{\rho_1}_{-\ii \tau_1}\cdots\h K^{\rho_n}_{-\ii \tau_n}
	\h J^{\sigma_1}_{-\ii \tau_{n+1}} \cdots\h J^{\sigma_m}_{-\ii \tau_{n+m}} \h O(0)\right)}_{T,c}\, .
\end{split}
\end{equation}
All the thermodynamic coefficients are linear combination of correlators~(\ref{mycorrrest})
with an appropriate $\wK$, $\wJ$ and $\h O$ components choice, and such expressions can be considered
as ``Kubo formulae'' for the corresponding coefficients. From the Lorentz generators definition~(\ref{Lorentzgen}), it follows that the~(\ref{mycorrrest}) are given by means of correlators of the symmetric stress-energy tensor and the operator $\h O$. In this work, we work out the (\ref{mycorrrest}) by
using the imaginary time formalism. The shifted boost and angular momentum generators, here obtained 
by definition~(\ref{boostrot}) in the rest frame, are given by Eq.~(\ref{jshift})
\begin{equation*}
\wJ^{\mu\nu}_{-\I\tau}= \h {\sf T}((-\I\tau,\vec{0})) \wJ^{\mu\nu} \h {\sf T}^{-1}((-\I\tau,\vec{0}))
\end{equation*}
and, according to definition~(\ref{Lorentzgen}), arise from the spatial integration of the time evolved stress-energy tensor operator
\begin{equation}
\label{imagtime}
\wT^{\mu\nu}(\tau,\vec{x})=\h {\sf T}((-\I\tau,\vec{0})) \wT^{\mu\nu}(0,\vec{x}) \h {\sf T}^{-1}((-\I\tau,\vec{0})).
\end{equation}
Thus, the correlators~(\ref{mycorrrest}) are expressed as a linear combination of the following basic structure with suitable indices:
\begin{equation}
\begin{split}\label{basicstrut}
C_{\mu_1\nu_1|\cdots|\mu_n\nu_n| i_1\cdots i_n}\equiv& \int_0^{|\beta|} \frac{\di\tau_1\cdots\di\tau_n}{|\beta|^n}
\int \D^3x_1\cdots\D^3x_n\times\\
&\times  \mean{{\rm T}_\tau\left(\h{T}_{\mu_1\nu_1}(X_1)\cdots\h{T}_{\mu_n\nu_n}(X_n) \h O(0)\right)}\ped{$T$,c}\, x_i\cdots x_n,
\end{split}
\end{equation}
where $X_n=(\tau_n,\vec{x}_n)$.

We are now in a position to determine the expansion in thermal vorticity of conserved currents
including the axial chemical potential $\mu_A = \zeta_A T$. But before that, it should be pointed out
that any scalar or a pseudo-scalar operator, denoted as $\h\phi$ and $\h\phi\ped{A}$
in Table~\ref{tab:TransfProp}, does not have chiral corrections up to second order. Indeed, the correlator 
between $\h\phi$ or $\h\phi\ped{A}$ and $\wJ$ or $\wK$ simply vanish at first-order because the 
latter are both vectors under rotation. Besides, it can be readily seen again from
Table~\ref{tab:TransfProp}, that no second-order correlator can be formed with $\h K$ and $\h J$ 
that transform at the same time even under time reversal and odd under parity. Hence, there 
are no chiral-vortical corrections even at second order for scalar and pseudoscalar operators.

\subsection{Stress-energy tensor}

The expansion up to second order in thermal vorticity of the stress-energy tensor with axial
chemical potential features additional terms with respect to the case $\zeta_A =0$ ~\cite{Buzzegoli:2017cqy,Becattini:2015nva}. Indeed, three new chiral coefficients appear
at the first order in thermal vorticity, $\mathbb{A},\mathbb{W}_1$ and $\mathbb{W}_2$:
\begin{equation}
\begin{split}\label{setdecomp}
\mean{\wT^{\mu\nu}}=&\,\mathbb{A}\,\epsilon^{\mu\nu\kappa\lambda}\alpha_\kappa u_\lambda+\mathbb{W}_1 w^\mu u^\nu +\mathbb{W}_2 w^\nu u^\mu\\
&+(\rho-\alpha^2 U_\alpha -w^2 U_w)u^\mu u^\nu -(p-\alpha^2D_\alpha-w^2D_w)\Delta^{\mu\nu}\\
&+A\,\alpha^\mu\alpha^\nu+Ww^\mu w^\nu+G_1 u^\mu\gamma^\nu+G_2 u^\nu\gamma^\mu+\mathcal{O}(\varpi^3)
\end{split}
\end{equation}
which can be obtained as:
\begin{equation}\label{setcoeff}
\begin{split}
\mathbb{A}=\mycorr{\,\wK^{3}\,\frac{\wT^{12}-\wT^{21}}{2}}_T,\quad
\mathbb{W}_1=\mycorr{\,\wJ^{3}\,\wT^{30}}_T,\quad
\mathbb{W}_2=\mycorr{\,\wJ^{3}\,\wT^{03}}_T,
\end{split}
\end{equation}
while for non chiral coefficients we found~\cite{Buzzegoli:2017cqy}\footnote{Differently from~\cite{Buzzegoli:2017cqy} here the stress-energy tensor is taken to be the canonical 
Dirac stress-energy tensor, which is non-symmetric. This is why we have $G_1$ and $G_2$.}
\begin{equation*}
\begin{split}
U_\alpha&=\frac{1}{2}\mycorr{\,\wK^3\,\wK^3\,\wT^{00}}_T,\qquad\hphantom{G_2=-} U_w=\frac{1}{2}\mycorr{\,\wJ^3\,\wJ^3\,\wT^{00}}_T,\\
D_\alpha&=\frac{1}{2}\mycorr{\,\wK^3\,\wK^3\,\wT^{11}}_T-\frac{1}{3}\mycorr{\,\wK^1\,\wK^2\,\frac{\wT^{12}+\wT^{21}}{2}}_T,\\
D_w&=\frac{1}{2}\mycorr{\,\wJ^3\,\wJ^3\,\wT^{11}}_T-\frac{1}{3}\mycorr{\,\wJ^1\,\wJ^2\,\frac{\wT^{12}+\wT^{21}}{2}}_T,\\
A&=\mycorr{\,\wK^1\,\wK^2\,\frac{\wT^{12}+\wT^{21}}{2}}_T, \qquad  W=\mycorr{\,\wJ^1\,\wJ^2\,\frac{\wT^{12}+\wT^{21}}{2}}_T,\\
G_1&=-\frac{1}{2}\mycorr{\,\{\wK^1,\,\wJ^2\}\,\wT^{03}}_T,\, \qquad G_2=-\frac{1}{2}\mycorr{\,\{\wK^1,\,\wJ^2\}\,\wT^{30}}_T.
\end{split}
\end{equation*}
%
%
%
%
It is useful to express the previous coefficients in terms of the quantity, see Eq.~(\ref{basicstrut}),
\begin{equation*}
C_{\mu\nu|\alpha\beta| i}\apic{se}\equiv\int_0^{|\beta|} \frac{\D\tau}{|\beta|}  \int \D^3x
 \mean{{\rm T}_\tau\left(\h{T}_{\mu\nu}(X)\,\h{T}_{\alpha\beta}(0)\right)}\ped{$T$,c}\, x_i,
\end{equation*}
which will be computed in Section~\ref{sec:firstordercoeff} for the massless Dirac field. By using the definition~(\ref{Lorentzgen}), we obtain
\begin{equation}\label{setcoeff2}
\begin{split}
\mathbb{A}&=\frac{1}{2}\left(C_{00|21|3}\apic{se}-C_{00|12|3}\apic{se}\right),\\
\mathbb{W}_1&=C_{02|30|1}\apic{se}-C_{01|30|2}\apic{se},\\
\mathbb{W}_2&=C_{02|03|1}\apic{se}-C_{01|03|2}\apic{se}.
\end{split}
\end{equation}

The conservation equation for the mean value of the stress-energy tensor:
\begin{equation*}
\de_\mu\mean{\wT^{\mu\nu}}=\de_\mu\tr\left(\wrho\,\wT^{\mu\nu}\right)=\mean{\de_\mu\wT^{\mu\nu}}=0
\end{equation*}
implies that not all the coefficients in~(\ref{setdecomp}) are independent. Indeed, enforcing 
the continuity equation, the following constraints for second order coefficients are obtained~\cite{Buzzegoli:2017cqy}:
\begin{align*}
U_\alpha&=-|\beta|\frac{\partial}{\partial|\beta|}\big(D_\alpha+A\big)-\big(D_\alpha+A\big),\\
U_w&=-|\beta|\frac{\partial}{\partial|\beta|}\big(D_w+W\big)-D_w+2A-3W,\\
G_1+G_2&=2\big(D_\alpha+D_w\big)+A+|\beta|\frac{\partial}{\partial|\beta|}W+3W,
\end{align*}
where all derivatives are to be taken with fixed $\zeta=\mu|\beta|$ and $\zeta\ped{A}=\mu\ped{A}|\beta|$.
For the new first-order chiral coefficients we have:
\begin{equation}\label{setchiralrel}
-2\mathbb{A}=|\beta|\frac{\de \mathbb{W}_1}{\de|\beta|}+3\mathbb{W}_1+\mathbb{W}_2,
\end{equation}
meaning that only $\mathbb{W_1}$ and $\mathbb{W}_2$ are really independent. In this work,
however, we have computed all the chiral coefficients for a massless Dirac field using 
eqs.~(\ref{setcoeff}) and used the condition (\ref{setchiralrel}) as a consistency check.

\subsection{Currents}
\label{currents}

Consider now the conserved current $\wj\ped{V}^\mu$ related to the charge $\wQ$ (\ref{ConsCharges}) 
and its transformation properties in Table~\ref{tab:TransfProp}. The only non vanishing terms 
up to the second order in thermal vorticity expansion turn out to be:
\begin{equation}\label{vcurrdecomp}
\mean{\wj\ped{V}^\mu}=n\ped{V}\,u^\mu -\left(\alpha^2 N\apic{V}_\alpha+w^2 N\apic{V}_\omega\right)u^\mu+W\apic{V}w^\mu+G\apic{V}\gamma^\mu
+\mathcal{O}(\varpi^3),
\end{equation}
where~\cite{Buzzegoli:2017cqy}
\begin{equation}
\label{vcurrevencoeff}
N\apic{V}_\alpha=\frac{\mycorr{\,\wK^3\,\wK^3\,\wj^0\ped{V}}_T}{2},\quad
N\apic{V}_w=\frac{\mycorr{\,\wJ^3\,\wJ^3\,\wj^0\ped{V}}_T}{2},\quad
G\apic{V}=\frac{\mycorr{\,\{\wK^1,\,\wJ^2\}\,\wj^3\ped{V}}_T}{2},
\end{equation}
and the chiral coefficient $W^V$ reads:
\begin{equation}\label{CVEcoeff}
W\apic{V}=\mycorr{\,\wJ^{3}\,\wj^3\ped{V}\,}_T\, ,
\end{equation}
which is the conductivity of the so-called Chiral Vortical Effect (CVE). We can write 
$W\apic{V}$ by means of the general correlator:
\begin{equation*}
C_{\mu\nu|\alpha| i}\apic{V}\equiv\int_0^{|\beta|} \frac{\D\tau}{|\beta|}  \int \D^3x 
\mean{{\rm T}_\tau\left(\h{T}_{\mu\nu}(X)\,\wj_{\alpha}\apic{V}(0)\right)}\ped{$T$,c}\, x_i,
\end{equation*}
as:
\begin{equation}\label{CVEcoeff2}
W\apic{V}=C_{02|3|1}\apic{V}-C_{01|3|2}\apic{V}.
\end{equation}
Taking the divergence of~(\ref{vcurrdecomp}), one realizes that the mean vector current 
is conserved \mbox{$\de_\mu\mean{\wj\ped{V}^\mu}=0$} if the $W\apic{V}$ coefficient 
fulfills the relation:
\begin{equation}\label{CVERel}
|\beta|\frac{\de W\apic{V}}{\de |\beta|}+3W\apic{V}=0.
\end{equation}
For a massless spinor field, using dimensional analysis, $W\apic{V}$ must be proportional 
to $|\beta|^{-3}$ ensuring that the previous relation is always verified.

Similarly, taking into account its transformation properties, the conserved axial current 
$\wj\ped{A}^\mu$ has the following expansion:
\begin{equation}\label{acurrdecomp}
\mean{\wj\ped{A}^\mu}=n\ped{A}\,u^\mu -\left(\alpha^2 N_\alpha\apic{A}+w^2 N_\omega\apic{A}\right)u^\mu+W\apic{A}w^\mu+G\apic{A}\gamma^\mu
+\mathcal{O}(\varpi^3).
\end{equation}
The so-called Axial Vortical Effect (AVE) shows up in the coefficient $W\apic{A}$
\begin{equation*}
W\apic{A}=\mycorr{\,\wJ^{3}\,\wj^3\ped{A}\,}_T
\end{equation*}
and has been discussed in detail in~\cite{Buzzegoli:2017cqy}, see also \cite{Prokhorov:2018qhq} for complete non chiral
corrections derived from a Wigner function ansatz. This coefficient does not
vanish for $\mu_A = 0$ and it is thus independent of the anomaly (see discussion 
in Section~\ref{discussion}). On the other hand, the proper chiral terms are given 
by the coefficients:
\begin{equation}\label{acurrcoeff}
\begin{aligned}
n\ped{A}&= \mean{\,\wj^{0}\ped{A}\,}_T, &
N_\alpha\apic{A}&=\frac{1}{2} \mycorr{\,\wK^{3}\, \wK^{3}\,\wj^{0}\ped{A}\,}_T,\\
N_\omega\apic{A}&=\frac{1}{2} \mycorr{\,\wJ^{3}\, \wJ^{3}\,\wj^{0}\ped{A}\,}_T, &
G\apic{A}&=\frac{1}{2} \mycorr{\{\wK^{1},\,\wJ^{2}\,\}\,\wj^{3}\ped{A}\,}_T,
\end{aligned}
\end{equation}
and, in terms of the auxiliary correlators
\begin{equation*}
C_{\mu\nu|\gamma\delta|\alpha| ij}\apic{A}=\int_0^{|\beta|}\frac{\D\tau_1}{|\beta|} \int_0^{|\beta|} \frac{\D\tau_2}{|\beta|} \int  \D^3x  \int  \D^3y
\mean{{\rm T}_\tau\left(\h{T}_{\mu\nu}(X)\,\h{T}_{\gamma\delta}(Y)\,\wj_{\alpha}\apic{A}(0)\right)}\ped{$T$,c}\, x_i y_j
\end{equation*}
they can be written as:
\begin{equation}\label{acurrcoeff2}
\begin{split}
N_\alpha\apic{A}&=\frac{1}{2}C_{00|00|0|33}\apic{A}, \\
N_w\apic{A}&=\frac{1}{2}\left(C_{01|01|0|22}\apic{A}-C_{01|02|0|21}\apic{A}-C_{02|01|0|12}\apic{A}+C_{02|02|0|11}\apic{A}\right),\\
G\apic{A}&=\frac{1}{2}\left(C_{00|03|3|11}\apic{A}-C_{00|01|3|13}\apic{A}+C_{03|00|3|11}\apic{A}-C_{01|00|3|31}\apic{A}\right).
\end{split}
\end{equation}
Again, conservation equation of the mean axial current entails a condition for $W\apic{A}$
\begin{equation}\label{AVERel}
|\beta|\frac{\de W\apic{A}}{\de |\beta|}+3W\apic{A}=0
\end{equation}
which is equivalent to that of the vector current and discussed in~\cite{Buzzegoli:2017cqy}.

\section{Discussion: \texorpdfstring{$\beta$}{beta} vs Landau frame in chiral hydrodynamics}

As an interlude preceding the explicit calculation of the thermodynamic coefficients, 
it is worth discussing the question of the frame, which is relevant for the hydrodynamics of chiral
fluids, that can be defined in general as fluids where the axial current is relevant. As has been
mentioned, the four-velocity in our formulae has been defined starting from the four-temperature 
vector $\beta$ which is a Killing vector in global equilibrium. This frame is called $\beta$ or 
thermodynamic frame \cite{Becattini:2014yxa,Kovtun2017} and, as it can be seen from Eq.~(\ref{setdecomp})
it does not coincide with Landau frame, where $u$ is by definition an eigenvector of the (symmetrized)
stress-energy tensor. However, unlike in the non-chiral case, where the difference between Landau frame
velocity and $\beta$ frame velocity shows up only at second order in the gradients of $u$ or $\beta$
owing to the terms in $u^\nu \gamma^\mu$~\cite{Becattini:2015nva}, see also~\cite{Ambrus:2017opa}
for an in-depth analysis; here the difference between 
the two velocities shows up at first order in the gradients because of the chiral terms in $u^\nu w^\mu$
in (\ref{setdecomp}) (see Appendix \ref{appa} for the transformation equations). 

This is especially important for chiral hydrodynamics because it implies that one cannot formulate
it at first order without taking these equilibrium non-dissipative terms in $u^\nu w^\mu$ into 
account. Even if one tried to eliminate them in the stress-energy tensor by going to the Landau frame,
they would reappear in the constitutive equations of the axial current at first order in the 
gradients; this is just what happens, see Appendix~\ref{appa}. For chiral-magneto-hydrodynamics, 
with dynamical electro-magnetic fields, the transition to Landau frame would be even more complicated 
because a change of the velocity definition implies a change of definition in the comoving electric 
field.

In conclusion, the $\beta$ frame appears as the most appropriate to write down hydrodynamic equations,
as advocated in \cite{Becattini:2014yxa,Kovtun2017}.

\section{Free massless Dirac field } 

We are now going to calculate the coefficients for the massless free Dirac field at finite temperature, 
vector and axial chemical potential. The basic quantity to compute is the grand-canonical partition 
function ${\mc Z}$:
\begin{equation*}
{\mc Z}=\tr\left[ \E^{-\beta (\h{H}-\mu \h{Q}-\mu\ped{A} \h{Q}\ped{A})}\right].
\end{equation*}
For the non interacting massless Dirac field described by the Lagrangian
\begin{equation*}
\mathcal{L}=\frac{\I}{2}\left(\bar\wPsi\gamma^\mu\partial_\mu\wPsi-(\partial_\mu\bar\wPsi)\gamma^\mu\wPsi\right)
\end{equation*}
the standard vector and axial currents read:
\begin{align*}
\h j\ped{V}^\mu&= \bar{\wPsi} \gamma^\mu \wPsi,\quad \hphantom{\gamma_5}\de_\mu \h j\ped{V}^\mu=0,\\
\h j\ped{A}^\mu&= \bar{\wPsi} \gamma^\mu \gamma^5 \wPsi,\quad \partial_\mu \h j\ped{A}^\mu=0,
\end{align*}
with the corresponding conserved vector and axial charges~(\ref{ConsCharges})
\begin{equation*}
\wQ\ped{V}=\int \D^3 x\, \h j\ped{V}^0(x),\qquad \wQ\ped{A}=\int \D^3 x\, \h j\ped{A}^0(x).
\end{equation*}
As it is known, the partition function can be written as a path integral of fields in
Euclidean spacetime
\begin{equation*}
{\mc Z}= C \int\ped{ABC} \mathcal{D} \bar{\Psi}\,\mathcal{D} \Psi \,\,
\exp \left\{ - \!\int_0^{|\beta|}\!\D\tau\!\int\!\D^3 x\,
\left[\frac{1}{2}\left(\bar\wPsi\tilde\gamma_\mu\partial_\mu\wPsi-\partial_\mu\bar\wPsi\tilde\gamma_\mu\wPsi\right) -\mu\ped{A} \bar\wPsi\tilde{\gamma}_0\gamma_5\wPsi \right] \right\}
\end{equation*}
with antiperiodic boundary condition (ABC) $\Psi(|\beta|,\vec{x})=-\Psi(0,\vec{x})$ and Euclidean 
gamma matrices $\tilde{\gamma}_\mu$, fulfilling the relation 
$\{\tilde \gamma_\mu ,\tilde \gamma_\nu \}=2 \delta_{\mu\nu}$. It is useful to define the following notation
\begin{equation}
\label{symbols}
P^\pm=(p_n\pm\ii\mu, {\bm p} ),\quad \quad X=(\tau ,{\bm x}), 
\quad \quad\sumint_{\{P\}}=\frac{1}{|\beta|}\sum_{n=-\infty}^\infty\int \frac{\di ^3 p }{(2\pi)^3},
\end{equation}
where $p_n=2\pi (n+1/2)/|\beta|$ with $n\in\mathbb{Z}$ are the Matsubara frequencies,
and to introduce the right and left chiral chemical potentials and the corresponding projection operator
\begin{equation*}
\mu\ped{R}=\mu+\mu\ped{A},\quad\mu\ped{L}=\mu-\mu\ped{A},\quad P\ped{R/L}^\pm=(\omega_n\pm\I \mu\ped{R/L},\vec{p}),
\quad \mathbb{P}_\chi=\frac{1+\chi\gamma_5}{2},
\end{equation*}
where $\chi=$R,L is assigned to be respectively $+1,-1$.
Then, the Dirac propagator in the imaginary time reads~\cite{Brown:1977eb}:
\begin{equation}\label{eq:Prop}
\mean{{\rm T}_\tau \wPsi_a(X)\bar{\wPsi}_b(Y)}_T=
\sum_\chi \sumint_{\{P\}} \frac{\e^{\ii P^+_\chi\cdot(X-Y)}}{{P^+_\chi}^2}\left(\mathbb{P}_\chi G(P^+_\chi)\right)_{ab},
\quad G(P)\equiv-\I\slashed{P}\, ,
\end{equation}
where latin characters $a,b,\dots$ denote spinorial indices, $\slashed P=\tilde\gamma_\mu P_\mu$ 
is the standard contraction between the (Euclidean) Dirac matrices $\tilde\gamma_\mu$ and the 
(Euclidean) four-momenta $P$.
The Euclidean canonical stress-energy tensor, see Eq.~(\ref{imagtime}), equals
\begin{equation*}
\h T_{\mu\nu}(X)=
\frac{\I^{\delta_{0\mu}+\delta_{0\nu}}}{2}
\bar{\wPsi}(X)\left[\tilde\gamma_\mu\oraw{\de}_\nu-\tilde\gamma_\mu\olaw{\de}_\nu\right]\wPsi(X),
\end{equation*}
where the $\I^{\delta_{0\mu}}$ factor stems from Wick rotation. The Belinfante-symmetrized 
stress-energy tensor used to construct all Poincaré generators, is simply the symmetrization 
of the previous one:
\begin{equation*}
\h T_{\mu\nu}(X)=
\frac{\I^{\delta_{0\mu}+\delta_{0\nu}}}{4}
\bar{\wPsi}(X)\left[\tilde\gamma_\mu\oraw{\de}_\nu-\tilde\gamma_\mu\olaw{\de}_\nu+\tilde\gamma_\nu\oraw{\de}_\mu-\tilde\gamma_\nu\olaw{\de}_\mu\right]\wPsi(X),
\end{equation*}
which can be expressed according to the point-splitting procedure as:
\begin{equation}\label{pointsplitset}
\h T_{\mu\nu}(X)= \lim_{X_1,X_2\to X} \mathcal{D}_{\mu\nu}(\de_{X_1},\de_{X_2})_{ab}\bar\wPsi (X_1)_a\wPsi (X_2)_b,
\end{equation}
where:
\begin{equation*}
\mathcal{D}_{\mu\nu} (\partial_{X_1},\partial_{X_2})=
\frac{\I^{\delta_{0\mu}+\delta_{0\nu}}}{4}\left[\tilde \gamma_\mu (\partial_{X_2}-\partial_{X_1})_\nu
+\tilde\gamma_\nu (\partial_{X_2}-\partial_{X_1})_\mu\right].
\end{equation*}
%

\subsection{First order correlators}\label{sec:firstordercoeff}

Here we outline the procedure to evaluate first and second order coefficients of a
general operator $\h{O}^{\alpha\dots\beta}(x)$ for free massless Dirac field theory.
We can write a generic correlator~(\ref{basicstrut}) related to first order corrections in vorticity as
\begin{equation}
\label{eq:firstorderprototype}
C_{\mu\nu|\alpha\dots\beta| i}\equiv\int_0^{|\beta|}   \frac{\D\tau}{|\beta|}  \int  \D^3x\,
\mean{{\rm T}_\tau\left(\h{T}_{\mu\nu}(X)\,\h{O}_{\alpha\dots\beta}(0)\right)}\ped{$T$,c}\, x_i,
\end{equation}
where $\wT_{\mu\nu}$ is the symmetric stress-energy tensor coming from Lorentz generators~(\ref{Lorentzgen})
and the two operators connected correlator is
\begin{equation*}
\mean{\h{A}\h{B}}\ped{$T$,c}= \mean{\h{A}\,\h{B}}_T- \mean{\h{A}}_T\, \mean{\h{B}}_T.
\end{equation*}

In the same way as stress-energy tensor in~(\ref{pointsplitset}),
we use the point-splitting procedure to express the operator $\h O$ as
\begin{equation*}
\h O_{\alpha\dots\beta}(Y)= \lim_{Y_1,Y_2\to Y} \mathcal{O}_{\alpha\dots\beta} (\de_{Y_1},\de_{Y_2})_{cd}\bar\wPsi (Y_1)_c\wPsi (Y_2)_d.
\end{equation*}
Thus, to compute~(\ref{eq:firstorderprototype}), we first consider
\begin{equation*}
\begin{split}
C(X,Y) &\equiv \mean{{\rm T}_\tau\left(\h{T}_{\mu\nu}(X)\,\h{O}_{\alpha\dots\beta}(Y)\right)}\ped{$T$,c} \\
& =\!\! \lim_{\substack{X_1 X_2\to X\\Y_1 Y_2\to Y}} {\Dpi_{\mu\nu}(\de_{X_1},\de_{X_2})}_{ab} {\oper(\de_{Y_1},\de_{Y_2})}_{cd}
\mean{{\rm T}_\tau\!\!\left( \bar{\wPsi}(X_1)_a\wPsi(X_2)_b \bar{\wPsi}(Y_1)_c\wPsi(Y_2)_d\right)}\ped{$T$,c}.
\end{split}
\end{equation*}
Now the connected correlator only concerns Dirac fields and can be computed using Wick's theorem
\begin{equation*}
\begin{split}
\mean{\bar{\psi}_1\psi_2\bar{\psi}_3\psi_4}\ped{c} & = \mean{\bar{\psi}_1\psi_2\bar{\psi}_3\psi_4}- \mean{\bar{\psi}_1\psi_2} \mean{\bar{\psi}_3\psi_4}\\
&= \mean{\bar{\psi}_1\psi_4} \mean{\psi_2\bar{\psi}_3}+ \mean{\bar{\psi}_1\psi_2} \mean{\bar{\psi}_3\psi_4}- \mean{\bar{\psi}_1\psi_2} \mean{\bar{\psi}_3\psi_4}\\
& = \mean{\bar{\psi}_1\psi_4} \mean{\psi_2\bar{\psi}_3}.
\end{split}
\end{equation*}
With this result and exchanging the order of the two anti-commuting fields $\bar{\wPsi}(X_1)_a\wPsi(Y_2)_d$
we are able to recreate a trace operation:
\begin{equation*}
\begin{split}
C(X,Y) &=    \lim_{\substack{X_1 X_2\to X\\Y_1 Y_2\to Y}}  {\Dpi_{\mu\nu}(\de_{X_1},\de_{X_2})}_{ab} {\oper(\de_{Y_1},\de_{Y_2})}_{cd}
\mean{{\rm T}_\tau \bar{\wPsi}(X_1)_a\wPsi(Y_2)_d} \mean{{\rm T}_\tau\wPsi(X_2)_b \bar{\wPsi}(Y_1)_c} \\
& =- \lim_{\substack{X_1 X_2\to X\\Y_1 Y_2\to Y}}\tr \left[ {\Dpi_{\mu\nu}(\de_{X_1},\de_{X_2})}  \mean{{\rm T}_\tau \wPsi(X_2) \bar{\wPsi}(Y_1)} {\oper(\de_{Y_1},\de_{Y_2})}
\mean{{\rm T}_\tau \wPsi(Y_2) \bar{\wPsi}(X_1)} \right].
\end{split}
\end{equation*}
Now, we express fermionic propagator in Fourier space as in~(\ref{eq:Prop})
\begin{gather*}
\mean{{\rm T}_\tau \wPsi(X_2)_a\bar{\wPsi}(Y_1)_b}=\sum_\chi \sumint_{\{P\}}\frac{\E^{\I P^+_\chi\cdot (X_2-Y_1)}}{{P^+_\chi}^2}\left(\mathbb{P}_\chi G(P^+_\chi)\right)_{ab} ,\\
\mean{{\rm T}_\tau \wPsi(Y_2)_a\bar{\wPsi}(X_1)_b}=\sum_{\chi'}\sumint_{\{Q\}}\frac{\E^{\I Q^+_{\chi'}\cdot (Y_2-X_1)}}{{Q^+_{\chi'}}^2}\left(\mathbb{P}_{\chi'} G(Q^+_{\chi'})\right)_{ab},
\end{gather*}
where, as previously, $P^+\ped{R/L}= (p_n+\I\mu\ped{R/L},\vec{p})$ and $Q^+\ped{R/L}=(q_m+\I\mu\ped{R/L},\vec{q})$.
The derivatives, appearing in the point-splitted operators $\Dpi$ and $\mc{O}$, act on the propagators and are readily obtained:
\begin{align*}
\de_X  \mean{{\rm T}_\tau \wPsi(X)_a\bar{\wPsi}(Y)_b}&= \sum_\chi \sumint_{\{P\}}\I P^+_\chi\E^{\I P^+_\chi\cdot (X-Y)}\frac{\left(\mathbb{P}_\chi G(P^+_\chi)\right)_{ab}}{{P^+_\chi}^2};\\
\de_Y  \mean{{\rm T}_\tau \wPsi(X)_a\bar{\wPsi}(Y)_b}&=\sum_\chi \sumint_{\{P\}}(-\I P^+_\chi)\E^{\I P^+_\chi\cdot (X-Y)}\frac{\left(\mathbb{P}_\chi G(P^+_\chi)\right)_{ab}}{{P^+_\chi}^2}.
\end{align*}
At this point, we take these derivatives, we send $X_1,X_2\to X$ and $Y_1,Y_2\to Y$ and we rename $Q$ into $-Q$; note that
$Q^+_{\chi'}=(q_m+\I\mu_{\chi'},\vec{q})$ goes to $-Q^-_{\chi'}=(\I\mu_{\chi'}-q_m,-\vec{q})$.
At the end, the correlator $C(X,Y)$ becomes
\begin{equation*}
C(X,Y) =  - \sum_\chi \sumint_{\{P,Q\}} \frac{\E^{\I (P^+_\chi +Q^-_{\chi'})(X-Y)}}{{P^+_\chi}^2\, {Q^-_\chi}^2}\,\,F_{\chi}(P^+,Q^-),
\end{equation*}
%
where we defined
\begin{equation*}
\begin{split}
F_{\chi}(P,Q)\equiv
\tr \left[ \mathbb{P}_\chi G(-Q_{\chi}) \Dpi_{\mu\nu}(\I Q_{\chi},\I P_\chi) G(P_\chi) {\oper(-\I P_\chi,-\I Q_{\chi})} \right].
\end{split}
\end{equation*}
The trace is to be carried out over spinorial indices by using the Euclidean $\tilde{\gamma}$ 
matrices properties:
\begin{align*}
\tr\left( \tilde{\gamma}_\mu \tilde{\gamma}_\nu\right)&=4\,\delta_{\mu\nu}, \\
\tr\left( \tilde{\gamma}_{\mu_{1}} \dots \tilde{\gamma}_{\mu_{2n+1}}\right)&=0, \\
\tr\left(\tilde{\gamma}_k\tilde{\gamma}_\lambda\tilde{\gamma}_\mu\tilde{\gamma}_\nu\gamma_5\right)&=4\epsilon^{k\lambda\mu\nu},\\
\tr\left( \tilde{\gamma}_k \tilde{\gamma}_\lambda \tilde{\gamma}_\mu \tilde{\gamma}_\nu\right)&=4\, \delta_{k\lambda}\delta_{\mu\nu}-4\, \delta_{k\mu}\delta_{\lambda\nu}+4\, \delta_{k\nu}\delta_{\lambda\mu}.
\end{align*}
However, all the operators $\h{O}$ that will be considered are analytic functions in the four-momentum 
$Q$ and $P$; as a result, the function $F_{\chi}$ will also be analytic in those variables. This
enables us to analytically entail the sum over the Matsubara frequencies.

Before carrying out the thermal sum, we send $Y$ to $0$, as dictated by~(\ref{eq:firstorderprototype}), 
and separate the momentum integral from the frequency sum in $\sumint$~(\ref{symbols})
\begin{equation*}
C(\tau,\vec{x})=C(X,0)=(-1)\sum_{\chi} \int \frac{\D^3 p}{(2\pi)^3}\int \frac{\D^3q}{(2\pi)^3} \E^{\I(\vec{p}+\vec{q})\cdot \vec{x}} \,S_{\chi}(\vec{p},\vec{q},\tau),
\end{equation*}
where we have defined
\begin{equation*}
S_{\chi}(\vec{p},\vec{q},\tau)\equiv \frac{1}{|\beta|}\sum_{n=-\infty}^\infty \frac{1}{|\beta|}
\sum_{m=-\infty}^\infty\frac{\E^{\I (p_n+\I\mu_\chi)\tau}}{(p_n+\I\mu_\chi)^2 +p^2}
\frac{\E^{\I (q_m -\I\mu_{\chi})\tau}}{(q_m-\I\mu_{\chi})^2 +q^2} F_{\chi}(P^+,Q^-).
\end{equation*}
As stated previously, $F_\chi$ is an analytic function, hence the sum over frequencies can be carried 
out using the formula:
\begin{equation}
\label{freqsum}
\frac{1}{|\beta|}\sum_{\{\omega_n\}} \frac{( \omega_n\pm\I\mu)^k \E^{\I (\omega_n\pm\I\mu) \tau}}{(\omega_n\pm\I\mu)^2 +\omega^2} =
\frac{1}{2\omega}\left[ (\I \omega)^k \E^{-\omega \tau}(1-n\ped{F}(\omega\mp\mu))-(-\I \omega)^k\E^{\omega \tau}n\ped{F}(\omega\pm\mu)\right],
\end{equation}
where $\omega$ could be the modulus of $p^2$ or of $q^2$ and $n\ped{F}$ is the Fermi-Dirac distribution function:
\begin{equation*}
n\ped{F}(\omega)=\frac{1}{\E^{|\beta|\omega}+1}.
\end{equation*}
Introducing the notation
\begin{equation}
\label{Ptilde}
\tilde{P}(\pm)=(\pm\I p,\vec{p}),\quad p^\pm_\chi= p\pm\mu_\chi
\end{equation}
and similarly for $\tilde{Q}(\pm)$, and taking advantage of Eq.~(\ref{freqsum}), we have
\begin{equation*}
\begin{split}
S_{\chi}(\vec{p},\vec{q},\tau) =\frac{1}{4 p q} \Bigl\{& F_{\chi}\left(\tilde{P}(+),\tilde{Q}(+) \right) \E^{-(p^-_\chi +q^+_{\chi})\tau}\big[1-n\ped{F}(p-\mu_\chi)\big]\big[1-n\ped{F}(q+\mu_{\chi})\big]+ \\
&- F_{\chi}\left(\tilde{P}(+),\tilde{Q}(-) \right) \E^{(-p^-_\chi +q^-_{\chi})\tau}\big[1-n\ped{F}(p-\mu_\chi)\big]n\ped{F}(q-\mu_{\chi})+ \\
&- F_{\chi}\left(\tilde{P}(-),\tilde{Q}(+) \right)\E^{(p^+_\chi -q^+_{\chi})\tau} n\ped{F}(p+\mu_\chi) \big[1-n\ped{F}(q+\mu_{\chi})\big]+ \\
&+F_{\chi}\left(\tilde{P}(-),\tilde{Q}(-) \right) \E^{(p^+_\chi +q^-_{\chi})\tau} n\ped{F}(p+\mu_\chi)  n\ped{F}(q-\mu_{\chi}) \Bigr\}.
\end{split}
\end{equation*}
Note that we have added and subtracted the chemical potential in the exponential in such a way 
that the corresponding energy couples with the relating argument of the Fermi distribution function 
$n\ped{F}$. This feature will be used later on.

Finally, putting together and integrating by parts, we can write the coefficient~(\ref{eq:firstorderprototype}) as:
\begin{equation*}
\begin{split}
C_{\mu\nu\alpha\dots\beta i} & =\int_0^{|\beta|}\frac{\D\tau}{|\beta|} \int\D^3 x\, x_i\, C(\tau,\vec{x}) \\
&=-\I\sum_{\chi}\int_0^{|\beta|}\frac{\D\tau}{|\beta|} \int \frac{\D^3p}{(2\pi)^3} \left[\frac{\de S_{\chi}(\vec{p},\vec{q},\tau)}{\de q^i}\right]_{\vec{q}=-\vec{p}},
\end{split}
\end{equation*}
that is an integral over time and momenta of a known function.
We can simplify this expression further with the following steps:
\begin{itemize}
	\item after the evaluation of $F_{\chi}(P,Q)$ calculate $S_{\chi}$;
	\item then derive $S_{\chi}(\vec{p},\vec{q},\tau)$ and replace $\vec{q}\to-\vec{p}$;
	\item integrate over solid angle $\Omega$: $\D^3 p=p^2\D p\,\D\Omega$ first and over $\tau$ thereafter;
	\item express all exponential factors with the Fermi distribution function using
	\begin{equation*}
		\E^{|\beta| E}=\frac{1}{n\ped{F}(E)}-1,\quad E=p\pm\mu\ped{R,L};
	\end{equation*}
	\item express the derivative of $n\ped{F}(E)$ in terms of its powers
	\begin{align*}
		n\ped{F}'(E)&=-|\beta| n\ped{F}^2(E) \E^{|\beta| E}=-|\beta| n\ped{F}(E)+|\beta| n\ped{F}^2(E),\\
		n\ped{F}''(E)&=|\beta|^2 n\ped{F}(E)-3|\beta|^2 n\ped{F}^2(E)+2 |\beta|^2 n\ped{F}^3(E),\\
		n\ped{F}'''(E)&=6|\beta|^3 n\ped{F}^4(E)-12|\beta|^3 n\ped{F}^3(E)+7|\beta|^3 n\ped{F}^2(E)-|\beta|^3 n\ped{F}(E);
	\end{align*}
	\item after the due simplifications, express powers of $n\ped{F}(E)$ in terms of its derivative:
	\begin{align*}
		n\ped{F}^4(E)&=\frac{1}{6|\beta|^3}n\ped{F}'''(E) +2 n\ped{F}^3(E)-\frac{7}{6}n\ped{F}^2(E)+\frac{1}{6}n\ped{F}(E),\\
		n\ped{F}^3(E)&=\frac{1}{2|\beta|^2}n\ped{F}''(E)+\frac{3}{2} n\ped{F}^2(E)-\frac{1}{2}n\ped{F}(E),\\
		n\ped{F}^2(E)&=\frac{1}{|\beta|}n\ped{F}'(E)+n\ped{F}(E),
	\end{align*}
	in this way a simple expression made only of $n\ped{F}(E)$ and its derivative is obtained;
	\item finally, integrate by part $n\ped{F}^{(k)}(E)$ when possible;
	\item sum up all the pieces with different $\chi$.
\end{itemize}
This procedure always leads to expressions of the form
\begin{equation}\label{GenCoeff}
\begin{split}
C(|\beta|,\mu\ped{R},\mu\ped{L})=K\int_0^\infty P_N(p)
\Big[ &n\ped{F}(p-\mu\ped{R})+\eta\,\eta\ped{A}\, n\ped{F}(p+\mu\ped{R})\\
&+\eta\ped{A}\, n\ped{F}(p-\mu\ped{L})+\eta\, n\ped{F}(p+\mu\ped{L})\Big]\D p,
\end{split}
\end{equation}
where $P_N$ is a polynomial of $N$ degree, $\eta$ is $+1$ if the correlator $C$ is even under charge transformation, otherwise it is $-1$, likewise  $\eta\ped{A}=\pm$ reflect parity transformation, and $K$ is a numerical factor. The correlators in the form~(\ref{GenCoeff}) can be easily integrated once we know the polynomial $P_N$.

The coefficients evaluated in~\cite{Buzzegoli:2017cqy} have either $\eta=+,\eta\ped{A}=+$ or $\eta=-,\eta\ped{A}=+$ and $\mu\ped{A}=0$, that means $\mu\ped{R}=\mu\ped{L}=\mu$ and thus have the form
\begin{equation*}
C(|\beta|,\mu)=2K\int_0^\infty P_\nu(p)
\left[ n\ped{F}^{(k)}(p-\mu)\pm\, n\ped{F}^{(k)}(p+\mu)\right]\D p.
\end{equation*}
So we can immediately obtain the dependence on chiral chemical potential for non chiral coefficients putting them in the form~(\ref{GenCoeff}) and setting
$K=\tilde{K}/2$, where $\tilde{K}$ is the numerical factor  evaluated in~\cite{Buzzegoli:2017cqy}.
The results obtained using this procedure are written in Section~\ref{sec:Results}.

\subsection{Second order correlators}

A second order correlator of the perturbative expansion~(\ref{basicstrut}) is the mean value 
of the local operator $\h O_{\alpha\dots\beta}$ with two Lorentz generators. We will follow
similar steps as those used for the first order correlators.
We start by writing the second order generic correlator~(\ref{basicstrut}) in Euclidean space-time as
\begin{equation}
\label{eq:secondorderprototype}
C_{\mu\nu|\gamma\delta|\alpha\dots\beta| ij}=\!\int_0^{|\beta|}\! \frac{\D\tau_1}{|\beta|}\! \int_0^{|\beta|}\! \frac{\D\tau_2}{|\beta|}\! \int\D^3x\!  \int\!\D^3y\,
\mean{{\rm T}_\tau \left(\h{T}_{\mu\nu}(X)\,\h{T}_{\gamma\delta}(Y)\,\h{O}_{\alpha\dots\beta}(0)\right)}\ped{$T$,c}\, x_i y_j,
\end{equation}
where $X=(\tau_1,\vec{x})$, $Y=(\tau_2,\vec{y})$ and the connected correlator is given by
\begin{equation*}
\mean{\h{A}\,\h{B}\,\h{C}}\ped{c}=\mean{\h{A}\,\h{B}\,\h{C}} -\mean{\h{A}}\mean{\h{B}\,\h{C}}
-\mean{\h{B}}\mean{\h{A}\,\h{C}} -\mean{\h{C}}\mean{\h{A}\,\h{B}} +2\mean{\h{A}}\mean{\h{B}}\mean{\h{C}}.
\end{equation*}
Again, using the point splitting procedure and taking advantage of the Wick's theorem, we can 
split the mean value of~(\ref{eq:secondorderprototype}) in two parts
\begin{equation}
\label{threepointfunc}
C(X,Y,Z)\equiv\mean{{\rm T}_\tau \left(\h{T}_{\mu\nu}(X)\,\h{T}_{\gamma\delta}(Y)\,\h{O}_{\alpha\dots\beta}(Z)\right)}\ped{$T$,c}=C_1+C_2,
\end{equation}
where $Z=(\tau_2,\vec{z})$ and
\begin{align*}
C_1=&\lim_{\substack{X_1 X_2\to X\\ Y_1 Y_2\to Y \\Z_1 Z_2\to Z}} (-1) \tr\left[\Dpi_{\mu\nu}(\de_{X_1},\de_{X_2})
	\mean{{\rm T}_\tau \wPsi(X_2) \bar{\wPsi}(Z_1)}\oper(\de_{Z_1},\de_{Z_2})\times\right. \\
	&\hphantom{\lim_{\substack{X_1 X_2\to X\\ Y_1 Y_2\to Y \\Z_1 Z_2\to Z}} (-1) \tr\Big[}\left.\times \mean{{\rm T}_\tau \wPsi(Z_2) \bar{\wPsi}(Y_1)} \Dpi_{\gamma\delta}(\de_{Y_1},\de_{Y_2})\mean{{\rm T}_\tau \wPsi(Y_2) \bar{\wPsi}(X_1)} \right],\\
C_2=&\lim_{\substack{X_1 X_2\to X\\ Y_1 Y_2\to Y \\Z_1 Z_2\to Z}}(-1) \tr\left[\Dpi_{\mu\nu}(\de_{X_1},\de_{X_2})
	\mean{{\rm T}_\tau \wPsi(X_2) \bar{\wPsi}(Y_1)}\Dpi_{\gamma\delta}(\de_{Y_1},\de_{Y_2})\times\right. \\
	&\hphantom{\lim_{\substack{X_1 X_2\to X\\ Y_1 Y_2\to Y \\Z_1 Z_2\to Z}} (-1) \tr\Big[}\left.\times \mean{{\rm T}_\tau \wPsi(Y_2) \bar{\wPsi}(Z_1)}\oper(\de_{Z_1},\de_{Z_2})\mean{{\rm T}_\tau \wPsi(Z_2) \bar{\wPsi}(X_1)} \right].
\end{align*}
Following similar procedure of first order correlators, we express the two point function in Fourier space~(\ref{eq:Prop})
and we can show that $C_1$ and $C_2$ are equal to 
\begin{equation*}
\begin{split}
C_1&= (-1)\sum_{\chi}\sumint_{\{P,Q,K\}}
	\frac{\E^{\I (P^+_{\chi} +K^-_{\chi})X} \E^{\I (Q^-_{\chi} -K^-_{\chi})Y} \E^{-\I (P^+_{\chi} +Q^-_{\chi})Z}}{{P^+_{\chi}}^2 {Q^-_{\chi}}^2 {K^-_{\chi}}^2}\,
	F_{1\,\chi}(P,Q,K),\\
C_2&=(-1)\sum_{\chi}\sumint_{\{P,Q,K\}}
	\frac{\E^{\I (P^-_{\chi} +K^+_{\chi})X} \E^{\I (Q^+_{\chi} -K^+_{\chi})Y} \E^{-\I (P^-_{\chi} +Q^+_{\chi})Z}}{{P^-_{\chi}}^2{Q^+_{\chi}}^2{K^+_{\chi}}^2}
	F_{2\,\chi}(P,Q,K),
\end{split}
\end{equation*}
where we denoted the momenta as $P^+_\chi= (p_n+\I\mu_\chi,\vec{p})$, $Q^+_{\chi}=(q_m+\I\mu_{\chi},\vec{q})$,\\ \mbox{$K^+_{\chi}=(k_l+\I\mu_{\chi},\vec{k})$},
and we defined the trace functions as:
\begin{equation*}
\begin{split}
F_{1\,\chi}(P,Q,K) =&\tr \Big[ \mathbb{P}_{\chi} G(-Q^-_{\chi})\Dpi_{\gamma\delta}(\I Q^-_{\chi} ,-\I K^-_{\chi})
	G(-K^-_{\chi})\Dpi_{\mu\nu}(\I K^-_{\chi} ,\I P^+_{\chi})\times \\
	&\hphantom{\tr \Big[}\times G(P^+_{\chi}) \oper(-\I P^+_{\chi},-\I Q^-_{\chi}) \Big],\\
F_{2\,\chi}(P,Q,K) =&\tr \Big[ \mathbb{P}_{\chi} G(-P^-_{\chi}) \Dpi_{\mu\nu}(\I P^-_{\chi},\I K^+_{\chi})
	G(K^+_{\chi}) \Dpi_{\gamma\delta}(-\I K^+_{\chi},\I Q^+_{\chi})\times \\
	&\hphantom{\tr \Big[}\times  G(Q^+_{\chi}) \oper(-\I Q^+_{\chi},-\I P^-_{\chi}) \Big].
\end{split}
\end{equation*}
Once the form of $\mc{O}$ is explicitly given, these two trace functions are computed thanks to the well known gamma matrix traces.
In general, we can say that as far as $\mc{O}$ is an analytic function of the derivative operator $\de_X$, the two
trace functions $F_1$ and $F_2$ are analytic functions on the four-momenta $P$, $Q$ and $K$.

Now, we send $Z$ to $0$ in Eq.~(\ref{threepointfunc}), as required by~(\ref{eq:secondorderprototype}),
and we separate the momenta integrals from the frequencies sums
\begin{equation}
\label{eq:C}
C(\tau_1,\tau_2,\vec{x},\vec{y})\equiv C(X,Y,0)=  -\sum_{\chi}\int \frac{\D^3p\, \D^3q\, \D^3k}{(2\pi)^9}\,
\E^{\I(\vec{p}+\vec{k})\cdot \vec{x}} \E^{\I(\vec{q}-\vec{k})\cdot \vec{y}} \, S_{\chi}(\vec{p},\vec{q},\vec{k},\tau_1,\tau_2),
\end{equation}
where we have defined
\begin{equation*}
\begin{split}
S_{\chi}(\vec{p},\vec{q},\vec{k},\tau_1,\tau_2)\! \equiv& \frac{1}{|\beta|^3}\!\!\!\sum_{n,m,l=-\infty}^\infty\!\!\Bigg[
\frac{\E^{\I (p_n+\I\mu_\chi) \tau_1+\I (q_m-\I\mu_{\chi}) \tau_2+\I (k_l-\I\mu_{\chi}) (\tau_1 -\tau_2)} F_{1\,\chi}(P^+,Q^-,K^-)}{[(p_n+\I\mu_\chi)^2 +p^2][(q_m-\I\mu_{\chi})^2 +q^2][(k_l-\I\mu_{\chi})^2 +k^2]}\\
&+\frac{\E^{\I (p_n-\I\mu_\chi) \tau_1+\I (q_m+\I\mu_{\chi}) \tau_2+\I (k_l+\I\mu_{\chi}) (\tau_1 -\tau_2)}F_{2\,\chi}(P^-,Q^+,K^+)}{[(p_n-\I\mu_\chi)^2 +p^2][(q_m+\I\mu_{\chi})^2 +q^2][(k_l+\I\mu_{\chi})^2 +k^2]}\Bigg].
\end{split}
\end{equation*}
Since $\tau_1$ and $\tau_2$ are always larger than zero in the integration~(\ref{eq:secondorderprototype}), the sums over $p_n$ and $q_m$
are made using~(\ref{freqsum}). For the $k_l$ sum, the result is still~(\ref{freqsum}) when $\tau=\tau_1-\tau_2>0$,
instead when $\tau=\tau_1-\tau_2<0$ the sum yields
\begin{equation*}
\frac{1}{|\beta|}\sum_{\{\omega_n\}} \frac{( \omega_n\pm\I\mu)^k \E^{\I (\omega_n\pm\I\mu) \tau}}{(\omega_n\pm\I\mu)^2 +\omega^2} =
\frac{1}{2\omega}\left[ (-\I \omega)^k\E^{\omega \tau}(1-n\ped{F}(\omega\pm\mu)) - (\I \omega)^k \E^{-\omega \tau}n\ped{F}(\omega\mp\mu)\right].
\end{equation*}
Reminding the definition~(\ref{Ptilde}) for $\tilde{P}(\pm)$, $\tilde{Q}(\pm)$, $\tilde{K}(\pm)$ and for $p^\pm_\chi$, $q^\pm_\chi$ and $k^\pm_\chi$,
after all the sums, we find
\begin{equation*}
\begin{split}
S_{\chi}(\vec{p},\vec{q},\vec{k},\tau_1,\tau_2) =\frac{1}{8\, p\, q\, k} \sum_{s_1,s_2,s_3=\pm}
\Bigl[& F_{1\,\chi}\left(\tilde{P}(s_1),\tilde{Q}(s_2),\tilde{K}(s_3)\right)S_{1,\chi}\\
&+ F_{2\,\chi}\left(\tilde{P}(s_1),\tilde{Q}(s_2),\tilde{K}(s_3)\right)\,S_{2\,\chi}\Bigr]
\end{split}
\end{equation*}
with
\begin{align*}
S_{1,\chi}\equiv& \exp\big[\!-s_1 p^{-s_1}_\chi\tau_1 \!-s_2 q^{s_2}_\chi\tau_2 \!-s_3 k^{s_3}_\chi(\tau_1\!-\tau_2)\big]
	\left[\theta(s_1)\!-n\ped{F}(p^{-s_1}_\chi) \right]\left[\theta(s_2)\!-n\ped{F}(q^{s_2}_\chi) \right]\times \\
	&\times \left[\left(\theta(s_3)\!-n\ped{F}(k^{s_3}_\chi)\right)\theta(\tau_1 \!- \tau_2)+\left(\theta(-s_3)\!-n\ped{F}(k^{s_3}_\chi)\right)\theta(\tau_2 \!- \tau_1)\right], \\
S_{2,\chi}\equiv&  \exp\big[\!-s_1 p^{s_1}_\chi\tau_1 \!-s_2 q^{-s_2}_\chi\tau_2 \!-s_3 k^{-s_3}_\chi(\tau_1\!-\tau_2)\big]
	\left[\theta(s_1)\!-n\ped{F}(p^{s_1}_\chi) \right]\left[\theta(s_2)\!-n\ped{F}(q^{-s_2}_\chi) \right]\times \\
	&\times \left[\left(\theta(s_3)\!-n\ped{F}(k^{-s_3}_\chi)\right)\theta(\tau_1 \!- \tau_2)+\left(\theta(-s_3)\!-n\ped{F}(k^{-s_3}_\chi)\right)\theta(\tau_2 \!- \tau_1)\right],
\end{align*}
where $\theta$ is the Heaviside theta function.
Now, we can take advantage of the formula
\begin{equation*}
\begin{split}
\int \D^3 x \int\D^3 y \,\, \E^{\I ({\bm p}+{\bm k} )\cdot {\bm x}}\E^{\I({\bm q}-{\bm k} )\cdot {\bm x} } 
x^i y^j=-(2\pi)^6 \frac{\de^2}{\partial k_i\de q_j}\delta^{(3)}(\bm p +\bm k)\delta^{(3)}(\bm q -\bm k)
\end{split}
\end{equation*}
to integrate over the coordinates $x$ and $y$ in the Eq.~(\ref{eq:secondorderprototype}) and, using Eq.~(\ref{eq:C}), we obtain:
\begin{equation*}
C_{\mu\nu|\gamma\delta|\alpha\dots\beta| ij}= \sum_{\chi}
\int_0^{|\beta|}\frac{\D\tau_1}{|\beta|} \int_0^{|\beta|}\frac{\D\tau_2}{|\beta|} \int \frac{\D^3p}{(2\pi)^3} \left[ \frac{\de^2 S_{\chi}(\vec{p},\vec{q},\vec{k},\tau_1,\tau_2)}{\de p^i \de q^j} \right]_{\vec{q}=-\vec{p},\vec{k}=-\vec{p}}.
\end{equation*}
From this point, we can adopt the same algorithm listed in the previous section, with an additional time integration,
to obtain simple expressions of the form~(\ref{GenCoeff}).

\subsection{Results}\label{sec:Results}
First, we consider both the canonical and symmetric stress energy tensor. In the Euclidean 
space-time they are given by
\begin{equation*}
\begin{split}
\h T_{\mu\nu}\apic{Can}(X)&=\frac{\I^{\delta_{0\mu}+\delta_{0\nu}}}{2}
	\bar{\wPsi}(X)\left[\tilde\gamma_\mu\oraw{\de}_\nu-\tilde\gamma_\mu\olaw{\de}_\nu\right]\wPsi(X),\\
\h T_{\mu\nu}\apic{Sym}(X)&=\frac{\I^{\delta_{0\mu}+\delta_{0\nu}}}{4}
\bar{\wPsi}(X)\left[\tilde\gamma_\mu\oraw{\de}_\nu-\tilde\gamma_\mu\olaw{\de}_\nu+\tilde\gamma_\nu\oraw{\de}_\mu-\tilde\gamma_\nu\olaw{\de}_\mu\right]\wPsi(X)
\end{split}
\end{equation*}
and they give different results for $\mathbb{A},\,\mathbb{W}_1,\,\mathbb{W}_2,\,G_1,\,G_2$, i.e. the coefficients related to the non symmetric part of the expansion
\begin{equation*}
\begin{split}
\mean{\wT^{\mu\nu}}=&\,\mathbb{A}\,\epsilon^{\mu\nu\kappa\lambda}\alpha_\kappa u_\lambda+\mathbb{W}_1 w^\mu u^\nu +\mathbb{W}_2 w^\nu u^\mu\\
&+(\rho-\alpha^2 U_\alpha -w^2 U_w)u^\mu u^\nu -(p-\alpha^2D_\alpha-w^2D_w)\Delta^{\mu\nu}\\
&+A\,\alpha^\mu\alpha^\nu+Ww^\mu w^\nu+G_1 u^\mu\gamma^\nu+G_2 u^\nu\gamma^\mu+\mathcal{O}(\varpi^3).
\end{split}
\end{equation*}
The non chiral coefficients can be ensued from~\cite{Buzzegoli:2017cqy} as described at the end of Section~\ref{sec:firstordercoeff} and are
\begin{equation*}
\begin{split}
\rho&= 3 p=\frac{30 \pi^2 \left(\zeta^2+\zeta\ped{A}^2\right)+15 \left(\zeta^4+6 \zeta^2 \zeta\ped{A}^2+\zeta\ped{A}^4\right)+7 \pi^4}{60 \pi^2 |\beta|^4},\\
U_\alpha&= 3 D_\alpha=\frac{3 \left(\zeta^2+\zeta\ped{A}^2\right)+\pi^2}{24 \pi^2 |\beta|^4},\\
G_1\apic{Sym}&= G_2\apic{Sym}=\frac{3 \left(\zeta^2+\zeta\ped{A}^2\right)+\pi^2}{18 \pi^2 |\beta|^4},\\
G_1\apic{Can} &=\frac{2 \left[3 \left(\zeta^2+\zeta\ped{A}^2\right)+\pi^2\right]}{9 \pi^2 |\beta|^4},\qquad
G_2\apic{Can} = -\frac{3 \left(\zeta^2+\zeta\ped{A}^2\right)+\pi^2}{9 \pi^2 |\beta|^4}.
\end{split}
\end{equation*}
All non-chiral coefficients reported in this section are also listed in Table~\ref{tab:EvenParity}.
%
\begin{table}[h!bt]
\small
\caption{The non-chiral thermodynamic coefficients up to second order in thermal vorticity
of the stress-energy tensor, electric and axial current for a free massless Dirac field (see 
Eqs.~(\ref{setdecomp}),(\ref{vcurrdecomp}),(\ref{acurrdecomp}) for definitions).
Here we use $T=1/|\beta|$, $\mu=\zeta\, T$ and $\mu\ped{A}=\zeta\ped{A}\, T$. }
\label{tab:EvenParity}
\newcommand\bstrut{\vphantom{\displaystyle\frac{\pi^{2^T}}{\sum}}}
\[
\begin{array}{||>{\displaystyle}c|>{\displaystyle}c||>{\displaystyle}c|>{\displaystyle}c||}
\hline
\rho=3p & \vphantom{\sum^{\Lambda^2}_{\Lambda^2}}\frac{7 \pi^2 T^4}{60}+\frac{\left(\mu^2+\mu\ped{A}^2\right) T^2}{2}+\frac{3 \mu^2 \mu\ped{A}^2}{2 \pi^2}+\frac{\mu^4+\mu\ped{A}^4}{4 \pi^2} &	A,\,W & 0\\
U_\alpha= 3 D_\alpha & \frac{T^4}{24}+\frac{\left(\mu^2+\mu\ped{A}^2\right) T^2}{8 \pi^2} &	n\ped{c} &	 \vphantom{\sum^{\Lambda^2}_{\Lambda^2}}\frac{\mu\,T^2}{3}+\frac{\mu\,\mu\ped{A}^2}{\pi^2}+\frac{\mu^3}{3 \pi^2}\\
U_w = 3 D_w & \vphantom{\sum^{\Lambda^2}_{\Lambda^2}}\frac{T^4}{8}+\frac{3 \left(\mu^2+\mu\ped{A}^2\right) T^2}{8 \pi^2} & N\apic{V}_\alpha & \frac{\mu\,  T^2}{4 \pi^2}\\
G_1\apic{Sym}= G_2\apic{Sym} & \frac{T^4}{18}+\frac{\left(\mu^2+\mu\ped{A}^2\right) T^2}{6 \pi^2} &	N\apic{V}_w & \vphantom{\sum^{\Lambda^2}_{\Lambda^2}}\frac{\mu\,  T^2}{4 \pi^2}\\
G_1\apic{Can} & \vphantom{\sum^{\Lambda^2}_{\Lambda^2}}\frac{2 T^4}{9}+\frac{2 \left(\mu^2+\mu\ped{A}^2\right) T^2}{3 \pi^2} & G\apic{V} & \frac{\mu\,  T^2}{6 \pi^2}\\
G_2\apic{Can} & -\frac{T^4}{9}-\frac{\left(\mu^2+\mu\ped{A}^2\right) T^2}{3 \pi^2} & W\apic{A} & \vphantom{\sum^{\Lambda^2}_{\Lambda^2}}\frac{T^3}{6}+\frac{\left(\mu^2+\mu\ped{A}^2\right) T}{2 \pi^2}\\
\hline
\end{array}
\]
\end{table}
%
Instead, the chiral coefficients are given by Eq.s~(\ref{setcoeff}), while accounting for~(\ref{setcoeff2})
can be calculated using the methods described above; they are found to be
\begin{equation}
\begin{split}\label{setchiralresults}
\mathbb{A}\apic{Sym} &= 0,\qquad\hphantom{+\zeta\ped{A}^2,\,\quad\mathbb{W}_2\apic{Can}= } \mathbb{A}\apic{Can}=\frac{\zeta\ped{A}\left(\pi^2+3\zeta^2+\zeta\ped{A}^2\right)}{6\pi^2|\beta|^4},\\
\mathbb{W}_1\apic{Sym}&=\mathbb{W}_2\apic{Sym}=\frac{\zeta\ped{A}\left(\pi^2+3\zeta^2+\zeta\ped{A}^2\right)}{3\pi^2|\beta|^4},\\
\mathbb{W}_1\apic{Can} &= \frac{\zeta\ped{A}\left(\pi^2+3\zeta^2+\zeta\ped{A}^2\right)}{2\pi^2|\beta|^4},\quad
\mathbb{W}_2\apic{Can}=  \frac{\zeta\ped{A}\left(\pi^2+3\zeta^2+\zeta\ped{A}^2\right)}{6\pi^2|\beta|^4}.\\
\end{split}
\end{equation}
We can readily check that the coefficients~(\ref{setchiralresults}) satisfy the relation~(\ref{setchiralrel}).
All chiral coefficients reported in this section are also listed in Table~\ref{tab:OddParity}.
%
\begin{table}[tb]
\small
\caption{The chiral thermodynamic coefficients up to second order in thermal vorticity
of the stress-energy tensor, electric and axial current for a free massless Dirac field 
(see Eqs.~(\ref{setdecomp}),(\ref{vcurrdecomp}),(\ref{acurrdecomp}) for definitions).
Here we use $T=1/|\beta|$, $\mu=\zeta\, T$ and $\mu\ped{A}=\zeta\ped{A}\, T$.}
\label{tab:OddParity}
\newcommand\bstrut{\vphantom{\displaystyle\frac{\pi^{2^T}}{\sum}}}
\[
\begin{array}{||>{\displaystyle}c|>{\displaystyle}c||>{\displaystyle}c|>{\displaystyle}c||}
\hline
\mathbb{A}\apic{Sym} & \vphantom{\sum^{\Lambda^2}_{\Lambda^2}}0 & W\apic{V} & \frac{\mu\, \mu\,\ped{A} T}{\pi^2 }\\
\mathbb{A}\apic{Can} & \frac{\mu\ped{A}\,T^3}{6}+\frac{\mu\ped{A}\,\mu^2\, T}{2\pi^2}+\frac{\mu\ped{A}^3\,T}{6\pi^2} &	\,\,n\ped{A}\,\, & \vphantom{\sum^{\Lambda^2}_{\Lambda^2}}\frac{\mu\ped{A}\,T^2}{3}+\frac{\mu\ped{A}\,\mu^2}{\pi^2}+\frac{\mu\ped{A}^3}{3\pi^2}\\
\mathbb{W}_1\apic{Sym}=\mathbb{W}_2\apic{Sym} &	\vphantom{\sum^{\Lambda^2}_{\Lambda^2}}\frac{\mu\ped{A}\,T^3}{3}+\frac{\mu\ped{A}\,\mu^2\, T}{\pi^2}+\frac{\mu\ped{A}^3\,T}{3\pi^2} &	N\apic{A}_\alpha & \frac{\mu\ped{A} T^2}{4 \pi^2 }\\
\mathbb{W}_1\apic{Can} & \frac{\mu\ped{A}\,T^3}{2}+\frac{3\mu\ped{A}\,\mu^2\, T}{2\pi^2}+\frac{\mu\ped{A}^3\,T}{2\pi^2} & N\apic{A}_w & \vphantom{\sum^{\Lambda^2}_{\Lambda^2}}\frac{\mu\ped{A} T^2}{4 \pi^2 }\\
\mathbb{W}_2\apic{Can} & \vphantom{\sum^{\Lambda^2}_{\Lambda^2}}\frac{\mu\ped{A}\,T^3}{6}+\frac{\mu\ped{A}\,\mu^2\, T}{2\pi^2}+\frac{\mu\ped{A}^3\,T}{6\pi^2} & G\apic{A} & \frac{\mu\ped{A} T^2 }{6 \pi^2}\\
\hline
\end{array}
\]
\end{table}

The Euclidean version of the Dirac field electric vector current $\wj\ped{V}^\mu=\bar\wPsi\gamma^\mu\wPsi$ is
\begin{equation*}
\wj\ped{V}^\mu=(-\I)^{1-\delta_{0\mu}}\bar\wPsi\tilde\gamma_\mu\wPsi.
\end{equation*}
The non chiral coefficients of the decomposition
\begin{equation*}
\mean{\wj\ped{V}^\mu}=n\ped{V}\,u^\mu-\left(\alpha^2 N\apic{V}_\alpha+w^2 N\apic{V}_\omega\right)u^\mu+W\apic{V}w^\mu+G\apic{V}\gamma^\mu
+\mathcal{O}(\varpi^3)
\end{equation*}
are obtained using the Eq.s~(\ref{vcurrevencoeff})
\begin{equation*}
\begin{aligned}
n\ped{V} &= \frac{\zeta  \left(\zeta^2+3 \zeta\ped{A}^2+\pi^2\right)}{3 \pi^2 |\beta|^3},&
N\apic{V}_\alpha &= \frac{\zeta }{4 \pi^2 |\beta|^3},\\
N\apic{V}_w &= \frac{\zeta }{4 \pi^2 |\beta|^3},&
G\apic{V} &= \frac{\zeta }{6 \pi^2 |\beta|^3},
\end{aligned}
\end{equation*}
while for the CVE conductivity $W\apic{V}$~(\ref{CVEcoeff}) evaluating the correlators in Eq.~(\ref{CVEcoeff2}) we obtain the well-known result
\begin{equation}\label{CVE}
	W\apic{V} = \frac{\zeta \zeta\ped{A}}{\pi^2 |\beta|^3}.
\end{equation}
The same is done for the axial current
\begin{equation*}
\wj\ped{A}^\mu=(-\I)^{1-\delta_{0\mu}}\bar\wPsi\tilde\gamma_\mu\gamma_5\wPsi,
\end{equation*}
which has similar expansion on thermal vorticity
\begin{equation*}
\mean{\wj\ped{A}^\mu}=n\ped{A}\,u^\mu-\left(\alpha^2 N_\alpha\apic{A}+w^2 N_\omega\apic{A}\right)u^\mu+W\apic{A}w^\mu+G\apic{A}\gamma^\mu
+\mathcal{O}(\varpi^3).
\end{equation*}
The only non chiral coefficient is the axial vortical effect conductivity $W\apic{A}$, for which we have~\cite{Buzzegoli:2017cqy}
\begin{equation}\label{AVE}
W\apic{A}=\frac{3 \left(\zeta^2+\zeta\ped{A}^2\right)+\pi^2}{6 \pi^2 |\beta|^3}.
\end{equation}
The chiral coefficients are again evaluated accounting for the relations~(\ref{acurrcoeff}) and~(\ref{acurrcoeff2}) 
\begin{equation*}
\begin{aligned}
n\ped{A}&= \frac{\zeta\ped{A}(\pi^2+3\zeta^2+\zeta\ped{A}^2)}{3\pi^2|\beta|^3}, &
N_\alpha\apic{A}&=\frac{\zeta\ped{A} }{4 \pi^2 |\beta|^3},\\
N_\omega\apic{A}&=\frac{\zeta\ped{A} }{4 \pi^2 |\beta|^3}, &
G\apic{A}&=\frac{\zeta\ped{A} }{6 \pi^2 |\beta|^3}.
\end{aligned}
\end{equation*}
As expected for dimensional analysis, the coefficients $W\apic{V}$~(\ref{CVE}) and $W\apic{A}$~(\ref{AVE}) respectively fulfill the relations~(\ref{CVERel}) and~(\ref{AVERel}).

\section{Discussion and summary}
\label{discussion}

Many of the coefficients presented here have been obtained with different methods.
The expectation values of the stress-energy tensor and the currents for free massless
Dirac field evaluated here are in agreement with the exact results in the case of pure
rotation and vanishing chemical potentials of~\cite{Ambrus:2014uqa}.
The CVE and AVE have been discussed at length over the past decade 
(see~\cite{Kharzeev:2015znc,Landsteiner:2016led,Jensen:2012kj}), and several calculations of 
$W\apic{V}$~(\ref{CVE}) and $W\apic{A}$~(\ref{AVE}) were presented in literature. 

As it was mentioned in Section~\ref{currents}, the AVE does not vanish for $\zeta_A=
\mu_A/T =0$, as it is apparent from eq.~(\ref{AVE}). This means that the AVE does not 
need chiral imbalance to show up and persists for a perfectly chirally balanced system. 
We interpret this feature - at least in the form in eq.~(\ref{AVE}) - as an evidence
that AVE is not related to the anomalous axial current divergence. As it was found by 
A. Vilenkin \cite{Vilenkin:1979ui,Vilenkin:1980zv,Vilenkin:1980ft} and lately highlighted 
in ref.~\cite{Buzzegoli:2017cqy}, an axial current proportional to vorticity arises 
because it is allowed by the symmetry of the density operator (this was also recognized 
in \cite{Son:2009tf}) with rotation and, indeed, the very same expression (\ref{AVE}) 
is found for free massless fermions, without coupling to either external or dynamical
gauge fields which make up the anomalous divergence of the axial current.
We understand that this statement goes against the conventional interpretation in 
literature, and yet some other authors have lately cast doubts about the anomalous origin 
of the AVE~\cite{Kalaydzhyan:2014bfa,Flachi:2017vlp}. We hope that our viewpoint will contribute to clarify the point.

As far as the stress-energy tensor is concerned, the consequence of parity breaking induced
by $\zeta_A$, is an energy flux $q^\mu$ along the vorticity $w^\mu$, see Eq.~(\ref{setdecomp}):
\begin{equation*}
q^\mu=\Delta^\mu_{\,\,\rho}\,u_\sigma\,\mean{\wT^{\sigma\rho}}=\mathbb{W}_2\,w^\mu.
\end{equation*}
To our knowledge, this kind of coefficient was obtained first by~\cite{Vilenkin:1979ui} as 
an energy flux resulting from neutrinos emitted by a rotating black hole; indeed, our result 
of $\mathbb{W}_2\apic{Can}$ for a free Dirac field~(\ref{setchiralresults}) perfectly agrees 
with the one in~\cite{Vilenkin:1979ui}(there is a factor $\frac{1}{2}$ of a difference
because Vilenkin considered only left-handed neutrinos). Lately, the same result for this 
coefficient was obtained by~\cite{Landsteiner:2011iq,Landsteiner:2012kd} using holographic 
techniques, by~\cite{Chen:2015gta,Hidaka:2017auj,Abbasi:2017tea} in chiral kinetic theory and in~\cite{Landsteiner:2012kd,Chowdhury:2015pba} evaluating Kubo formulae in finite temperature 
field theory. 

It is also worth noting that thermodynamic equilibria with vorticity imply different mean
values for the canonical and symmetric stress-energy tensor~\cite{Becattini:2011ev,Becattini:2012pp}.
This is seen here for the coefficients $\mathbb{A},\,\mathbb{W}_1,\,\mathbb{W}_2,\,G_1$ and $G_2$.
Particularly, the coefficient $\mathbb{A}$ vanishes if the stress-energy tensor is symmetric
but not for the canonical. 

The second order coefficients $N_\alpha\apic{A},\,N_w\apic{A}$ and $G\apic{A}$, see 
eqs~(\ref{acurrdecomp}) and~(\ref{acurrcoeff}) are indeed newly obtained and they appear 
as quantum corrections of the axial current in presence of acceleration and rotation. They 
are the axial counterpart of the second-order equilibrium corrections of the vector current discussed in~\cite{Buzzegoli:2017cqy}.  
$N_\alpha\apic{A},\,N_w\apic{A}$ are corrections along the fluid velocity and hence modify 
the axial charge density, while $G\apic{A}$ yields a chiral flow along the four-vector 
$\gamma^\mu$ defined in Eq.~(\ref{transversedir}). 

To summarize, we have studied global thermodynamic equilibrium with acceleration 
and vorticity with axial charge chemical potential inducing a macroscopic parity 
breaking. We carried out an expansion of the stress-energy tensor at the second 
order in the thermal vorticity tensor, including acceleration and vorticity and
obtained the constitutive equations at equilibrium of the stress-energy tensor as
well as of the vector and axial currents. These equations may be phenomenologically 
relevant for the physics of the Quark Gluon Plasma and especially for its hydrodynamic
modelling in presence of chirality imbalance.

\acknowledgments

M. B. carried out part of this work while visiting Stony Brook University 
(New York, USA). We would like to thank C. Bonati, E. Grossi, D. Kharzeev for 
stimulating discussions on the subject matter.

\appendix

\section{From \texorpdfstring{$\beta$}{beta} to Landau Frame}\label{appa}
In this work we choose the fluid velocity $u$ in the direction of the Killing $\beta$ four-vector.
This choice corresponds to the $\beta$ or thermodynamic frame. While this is a natural frame for
the generalized equilibrium, this is not the most common choice in literature and it is not the
frame typically used in numerical codes. It is then of interest to translate the results of this
paper into the Landau frame.

To accomplish this task, it suffices to establish the relation between the $\beta$-frame velocity
$u$ and the Landau frame velocity $u\ped{L}$.
By definition, the Landau velocity $u\ped{L}$ is the eigenvector of the stress energy tensor:
\begin{equation*}
u_{{\rm L}\mu}T^{\mu\nu}=\rho\ped{L}u\ped{L}^{\nu}
\end{equation*}
where $\rho\ped{L}$ is both the eigenvalue and the energy density in the Landau frame.
Being a four-vector, the Landau velocity $u\ped{L}$ can be expressed in terms of the tetrad
$\{u,\alpha,w,\gamma\}$ defined in Eq.s (\ref{accvort}) and (\ref{transversedir}):
\begin{equation}
\label{Landauvel}
u\ped{L}^\mu=a\,u^\mu+b\,\frac{w^\mu}{|w|}+c\,\frac{\gamma^\mu}{|\gamma|}
  +d\,\frac{\alpha^\mu}{|\alpha|}\, ,
\end{equation}
where $|w|=\sqrt{-w^2},\,|\gamma|=\sqrt{-\gamma^2},\,|\alpha|=\sqrt{-\alpha^2}$
and $a,b,c$ and $d$ are four unknown constants such that $u\ped{L}^\mu u_{{\rm L}\mu}=1$, i.e.
\begin{equation*}
a^2-b^2-2bd\frac{\alpha\cdot w}{|\alpha||w|}-c^2-d^2=1.
\end{equation*}
Furthermore, since for a vanishing thermal vorticity we expect that the two thermodynamic frames
coincide, for $\varpi=0$ we must have that $a=1$ and $\rho\ped{L}=\rho$.
This means that the leading term of $a$ and $\rho\ped{L}$ is zeroth order in thermal vorticity.

Now, taking the stress-energy tensor mean value at the second order expansion on thermal vorticity
\begin{equation*}
\begin{split}
T^{\mu\nu}=&\,\mathbb{A}\,\epsilon^{\mu\nu\kappa\lambda}\alpha_\kappa u_\lambda+\mathbb{W}_1 w^\mu u^\nu +\mathbb{W}_2 w^\nu u^\mu\\
&+(\rho-\alpha^2 U_\alpha -w^2 U_w)u^\mu u^\nu -(p-\alpha^2D_\alpha-w^2D_w)\Delta^{\mu\nu}\\
&+A\,\alpha^\mu\alpha^\nu+Ww^\mu w^\nu+G_1 u^\mu\gamma^\nu+G_2 u^\nu\gamma^\mu+\mathcal{O}(\varpi^3)
\end{split}
\end{equation*}
and contracting it with the landau velocity (\ref{Landauvel}), we obtain
\begin{equation*}
\begin{split}
u_{{\rm L}\mu}T^{\mu\nu}=&
\left(a\,\rho\ped{eff}-b\,\mathbb{W}_1|w|+d\,\mathbb{W}_1\frac{\alpha\cdot w}{|\alpha|}\right)u^\nu\\
  &+\left(a\,\mathbb{W}_2-b\,p\ped{eff}\frac{1}{|w|}-b\,W|w|
  +c\,\mathbb{A}\frac{|\alpha|^2}{|\gamma|}+d\,W\frac{\alpha\cdot w}{|\alpha|}\right)w^\nu+\\
&+\left(a\,G_1-b\,\mathbb{A}\frac{1}{|w|}-c\,p\ped{eff}\frac{1}{|\gamma|}\right)\gamma^\nu+\\
&+\left(b\,A\frac{\alpha\cdot w}{|w|}+c\,\mathbb{A}\frac{\alpha\cdot w}{|\gamma|}
 -d\,p\ped{eff}\frac{1}{|\alpha|}-d\,A|\alpha|\right)\alpha^\nu+\mathcal{O}(\varpi^3),
\end{split}
\end{equation*}
where for simplicity we defined $\rho\ped{eff}\equiv\rho-\alpha^2 U_\alpha -w^2 U_w$ and
$p\ped{eff}\equiv p-\alpha^2D_\alpha-w^2D_w$.
Equating this expression to $\rho\ped{L}u\ped{L}^{\nu}$, with $u\ped{L}$ given by (\ref{Landauvel}),
we obtain the five equations in five unknown variables $(a,b,c,d,\rho\ped{L})$ that
diagonalize the stress energy tensor (at second order in thermal vorticity):
\begin{equation}
\label{eqslin}
\begin{split}
a\,\rho\ped{eff}-b\,\mathbb{W}_1|w|+d\,\mathbb{W}_1\frac{\alpha\cdot w}{|\alpha|} &= a\,\rho\ped{L}\\
a\,\mathbb{W}_2-b\,p\ped{eff}\frac{1}{|w|}-b\,W|w| +c\,\mathbb{A}\frac{|\alpha|^2}{|\gamma|}
  +d\,W\frac{\alpha\cdot w}{|\alpha|}&=b\,\rho\ped{L}\,\frac{1}{|w|}\\
a\,G_1-b\,\mathbb{A}\frac{1}{|w|}-c\,p\ped{eff}\frac{1}{|\gamma|}
   &=c\,\rho\ped{L}\frac{1}{|\gamma|}\\
b\,A\frac{\alpha\cdot w}{|w|}+c\,\mathbb{A}\frac{\alpha\cdot w}{|\gamma|}
 -d\,p\ped{eff}\frac{1}{|\alpha|}-d\,A|\alpha|&=d\,\rho\ped{L}\frac{1}{|\alpha|}\\
 a^2-b^2-2bd\frac{\alpha\cdot w}{|\alpha||w|}-c^2-d^2 &=1.
\end{split}
\end{equation}

To solve it, we can simplify the equations furtherer.
Indeed, we can write the first Eq. of (\ref{eqslin}) as
\begin{equation}
\label{eqforb}
\frac{b}{a}=\frac{\rho\ped{eff}-\rho\ped{L}}{\mathbb{W}_1}\frac{1}{|w|}
  +\frac{d}{a}\frac{\alpha\cdot w}{|\alpha||w|}.
\end{equation}
In the same way, we can write the third Eq. of (\ref{eqslin}) isolating $b/a$:
\begin{equation*}
\frac{b}{a}=\frac{G_1}{\mathbb{A}}|w|
 -\frac{c}{a}\frac{\rho\ped{L}+p\ped{eff}}{\mathbb{A}}\frac{|w|}{|\gamma|};
\end{equation*}
hence, equating the previous two, we obtain
\begin{equation}
\label{eqforc}
\frac{c}{a}=\frac{\mathbb{A}}{\rho\ped{L}+p\ped{eff}}\left[\frac{G_1}{A}|\gamma|+
 \frac{\rho\ped{eff}-\rho\ped{L}}{\mathbb{W}_1}\frac{|\gamma|}{|w|^2}
 +\frac{d}{a}\frac{(\alpha\cdot w)|\gamma|}{|\alpha||w|^2}\right].
\end{equation}
From Eq. (\ref{eqforc}) we conclude that $c$ is at least first order in $\varpi$.
While, from the fourth Eq. of (\ref{eqslin}) we have
\begin{equation}
\label{eqford}
d=\frac{\mathbb{A}}{\rho\ped{L}+p\ped{eff}+A|\alpha|^2}(b\,|\gamma|+c\,|w|)\frac{\alpha\cdot w}{|w||\gamma|}|\alpha|,
\end{equation}
showing that $d$ is at least second order in thermal vorticity.

Since we are interested in second order solution and $d$ is already second order,
the Eq.~(\ref{eqforb}) gives
\begin{equation}
\label{onlyb}
\frac{b}{a}=\frac{\rho\ped{eff}-\rho\ped{L}}{\mathbb{W}_1}\frac{1}{|w|}+\mathcal{O}(\varpi^3),
\end{equation}
while the Eq. (\ref{eqforc}) becomes
\begin{equation}
\label{onlyc}
\frac{c}{a}=\frac{\mathbb{A}}{\rho\ped{L}+p\ped{eff}}\left[\frac{G_1}{A}|\gamma|+
 \frac{\rho\ped{eff}-\rho\ped{L}}{\mathbb{W}_1}\frac{|\gamma|}{|w|^2}\right]+\mathcal{O}(\varpi^3).
\end{equation}
We can also note that in the second equation of (\ref{eqslin}) the term in $d$ is higher order
and can be omitted, as a consequence, after the replacement of Eq.s (\ref{onlyb}) and (\ref{onlyc}), the second equation of (\ref{eqslin}) becomes a third grade equation for $\rho\ped{L}$:
\begin{equation*}
\mathbb{W}_1 \mathbb{W}_2(\rho\ped{L}+p\ped{eff})|w|^2
+(\rho\ped{L}+p\ped{eff})(\rho\ped{L}-\rho\ped{eff})(\rho\ped{L}+p\ped{eff}+W|w|^2)
+\mathbb{W}_1 G_1 A |w|^2|\alpha|^2+\mathbb{A}(\rho\ped{eff}-\rho\ped{L})|\alpha|^2=0.
\end{equation*}
We find that the only solution of this Eq. reproducing $\rho\ped{L}=\rho$ for $\varpi=0$ is:
\begin{equation*}
\rho\ped{L}=\rho\ped{eff}-\frac{\mathbb{W}_1\mathbb{W}_2}{\rho+p}|w|^2+\mathcal{O}(\varpi^4).
\end{equation*}
Replacing this solution in (\ref{onlyb}), we obtain
\begin{equation*}
\frac{b}{a}=\frac{\mathbb{W}_2}{\rho+p}|w|+\mathcal{O}(\varpi^3)
\end{equation*}
and replacing it in (\ref{onlyc}) gives
\begin{equation*}
\frac{c}{a}=\frac{G_1(\rho+p)+\mathbb{A}\mathbb{W}_2}{(\rho+p)^2}|\gamma|+\mathcal{O}(\varpi^3).
\end{equation*}
From these two, we learn that $b$ is actually first order on thermal vorticity and $c$ is second
order, thus from (\ref{eqford}) it follows that $d$ is in fact third order and can be set to zero.
Consequently, the last equation to solve is $a^2-b^2-c^2=1$, that gives
\begin{equation*}
a=\frac{1}{\sqrt{1-(b/a)^2-(c/a)^2}}
=1+\frac{1}{2}\frac{\mathbb{W}_2^2}{(\rho+p)^2}|w|^2+\mathcal{O}(\varpi^3).
\end{equation*}

In conclusion, we found that the relation between the Landau fluid velocity $u\ped{L}$
and the $\beta$-frame fluid velocity $u$ is
\begin{equation*}
u\ped{L}=\left(1+\frac{1}{2}\frac{\mathbb{W}_2^2}{(\rho+p)^2}|w|^2\right)u+
\frac{\mathbb{W}_2}{\rho+p} w
+\frac{G_1(\rho+p)+\mathbb{A}\mathbb{W}_2}{(\rho+p)^2}\gamma+\mathcal{O}(\varpi^3)
\end{equation*}
or, reverting it, $u$ is given by 
\begin{equation}
\label{uLandau}
u=\left(1-\frac{1}{2}\frac{\mathbb{W}_2^2}{(\rho+p)^2}|w|^2\right)u\ped{L}
-\frac{\mathbb{W}_2}{\rho+p}w
-\frac{G_1(\rho+p)+\mathbb{A}\mathbb{W}_2}{(\rho+p)^2}\gamma+\mathcal{O}(\varpi^3).
\end{equation}
This transformation at first order in vorticity and the following relation between CVE conductivity
in Landau frame with $\mathbb{W}_2$ and $W\apic{V}$ were also pointed out in~\cite{Landsteiner:2012kd}.
At this point, taking advantage of (\ref{uLandau}), we can express the stress-energy tensor
mean value at second order in thermal vorticity (\ref{setdecomp}) in the Landau frame:
\begin{equation*}
\begin{split}
T^{\mu\nu}=&\,\mathbb{A}\epsilon^{\mu\nu\kappa\lambda}\alpha_\kappa u_{{\rm L}\lambda}
-\frac{\mathbb{A}\mathbb{W}_2}{\rho+p}\epsilon^{\mu\nu\kappa\lambda}\alpha_\kappa w_\lambda
+(\mathbb{W}_1-\mathbb{W}_2)u\ped{L}^\nu w^\mu - p\ped{eff}\,\Delta\ped{L}^{\mu\nu}\\
&+\left(\rho\ped{eff}-\frac{\mathbb{W}_2^2}{\rho+p}|w|^2\right)u\ped{L}^\mu u\ped{L}^\nu
+A\alpha^\mu \alpha^\nu+\frac{W(\rho+p)+\mathbb{W}_1\mathbb{W}_2}{\rho+p}w^\mu w^\nu\\
&+\frac{\mathbb{A}\mathbb{W}_2}{\rho+p}u\ped{L}^\mu\gamma^\nu
+\frac{(G_2-G_1)(\rho+p)+\mathbb{A}\mathbb{W}_2}{\rho+p}u\ped{L}^\nu\gamma^\mu.
\end{split}
\end{equation*}
Furthermore, using (\ref{uLandau}), we can also write the vectorial current (\ref{vcurrdecomp}) in the Landau frame
\begin{equation*}
\begin{split}
j\ped{V}^\mu=& \, n\ped{V}\left(1-\frac{1}{2}\frac{\mathbb{W}_2^2}{(\rho+p)^2}|w|^2\right) 
 u\ped{L}^\mu-\left(\alpha^2\,N\apic{V}_\alpha+w^2\,N\apic{V}_w\right)u\ped{L}^\mu\\
 &+\left(W\apic{V}-n\ped{V}\frac{\mathbb{W}_2}{\rho+p} \right)w^\mu
  +\left(G\apic{V}-n\ped{V}\frac{G_1(\rho+p)+\mathbb{A}\mathbb{W}_2}{(\rho+p)^2}\right)\gamma^\mu.
\end{split}
\end{equation*}
Instead, replacing (\ref{uLandau}) in (\ref{acurrdecomp}), we obtain the axial current in the Landau frame
\begin{equation*}
\begin{split}
j\ped{A}^\mu=& \, n\ped{A}\left(1-\frac{1}{2}\frac{\mathbb{W}_2^2}{(\rho+p)^2}|w|^2\right) 
 u\ped{L}^\mu-\left(\alpha^2\,N\apic{A}_\alpha+w^2\,N\apic{A}_w\right)u\ped{L}^\mu\\
 &+\left(W\apic{A}-n\ped{A}\frac{\mathbb{W}_2}{\rho+p}\right)w^\mu
  +\left(G\apic{A}-n\ped{A}\frac{G_1(\rho+p)+\mathbb{A}\mathbb{W}_2}{(\rho+p)^2}\right)\gamma^\mu.
  \end{split}
\end{equation*}
%



\end{document}